\documentclass[manuscript,screen,nonacm]{acmart}
\setcopyright{none}
\AtBeginDocument{%
  }

\usepackage{amsthm}

\newtheoremstyle{compact}
  {3pt}   
  {3pt}   
  {\normalfont}
  {}
  {\bfseries}
  {}
  {.5em}
  {}

\theoremstyle{compact}
\newtheorem{hyp}{}

\makeatletter
\@addtoreset{hypsub}{hyp}
\makeatother

\renewcommand\footnotetextcopyrightpermission[1]{}

\usepackage[capitalize]{cleveref}
\usepackage{xspace}
\usepackage{xcolor}
\usepackage{listings}

\lstnewenvironment{promptbox}[1][Prompt]
  {\lstset{
    basicstyle=\small\ttfamily,
    breaklines=true,
    breakatwhitespace=true,
    frame=single,
    framesep=8pt,
    rulecolor=\color{black!35},
    backgroundcolor=\color{black!5},
    columns=fullflexible,
    keepspaces=true,
    title={#1},
  }}
  {}

\newcommand{\OFull}{Openness\xspace}
\newcommand{\CFull}{Conscientiousness\xspace}
\newcommand{\EFull}{Extraversion\xspace}
\newcommand{\AFull}{Agreeableness\xspace}
\newcommand{\NFull}{Neuroticism\xspace}

\begin{document}

\title[Does Personality Persuade?]{Does Personality Persuade? Investigating Personality Trait Prompting and Persuasion in Large Language Models}
\subtitle{Preprint}

\newcommand{\ourinstitution}{Hasselt University - Flanders Make}
\newcommand{\ourdepartment}{Digital Future Lab}

\author{Ruben Boone}
\email{ruben.boone@uhasselt.be}
\authornote{Authors contributed equally to this research.}
\orcid{0009-0004-1743-3117}
\correspondingauthor
\affiliation{%
  \institution{\ourinstitution}
  \department{\ourdepartment}
  \city{Diepenbeek}
  \state{Limburg}
  \country{Belgium}
}

\author{Gilles Eerlings}
\orcid{0009-0005-7765-6774}
\email{gilles.eerlings@uhasselt.be}
\authornotemark[1]
\affiliation{%
  \institution{\ourinstitution}
  \department{\ourdepartment}
  \city{Diepenbeek}
  \state{Limburg}
  \country{Belgium}
}

\author{Sebe Vanbrabant}
\orcid{0009-0001-7996-6048}
\email{sebe.vanbrabant@uhasselt.be}
\authornotemark[1]
\affiliation{%
  \institution{\ourinstitution}
  \department{\ourdepartment}
  \city{Diepenbeek}
  \state{Limburg}
  \country{Belgium}
}

\author{Davy Vanacken}
\orcid{0000-0001-8436-5119}
\email{davy.vanacken@uhasselt.be}
\affiliation{%
  \institution{\ourinstitution}
  \department{\ourdepartment}
  \city{Diepenbeek}
  \state{Limburg}
  \country{Belgium}
}

\author{Kris Luyten}
\orcid{0000-0002-4194-1101}
\email{kris.luyten@uhasselt.be}
\affiliation{%
  \institution{\ourinstitution}
  \department{\ourdepartment}
  \city{Diepenbeek}
  \state{Limburg}
  \country{Belgium}
}

\renewcommand{\shortauthors}{Boone, Eerlings, Vanbrabant et al.}

\begin{abstract}
Large language models (LLMs) can be prompted to express different personality profiles, but it remains unclear how reliably they follow these prompts and whether the personality they express affects their persuasiveness. We address these questions in two evaluations. First, we assess Big Five personality prompting with the IPIP-NEO-120, comparing prompt formats across repeated generations and measuring trait control, cross-trait spillover, and unconditioned model profiles. Second, we conduct a within-subjects study in which participants discuss morally ambiguous decisions with LLM agents prompted to exhibit personality profiles aligned to the participant, mismatched, neutral, or unconditioned. Our calibration experiments show that prompted personality profiles can be highly reproducible and strongly responsive to requested trait levels, although manipulating one trait can also cause smaller changes in other traits. We further find that unconditioned LLMs do not constitute personality-neutral baselines: the models examined exhibit systematic and highly consistent personality profiles even without explicit personality instructions. In the user study (N=32), personality alignment did not significantly affect perceived persuasiveness, stance switching, or changes in decision confidence. Participants changed their initial stance in only 8 of 128 conversations, and we found no significant associations between their OCEAN personality traits and susceptibility to persuasion. However, higher self-reported AI persuasion literacy was associated with less frequent changes in decision confidence. An exploratory analysis further revealed that agents predominantly relied on logical appeals, with other persuasion strategies varying across conditions and dilemmas. Our findings highlight the importance of distinguishing between an LLM's ability to express a prompted personality and its actual influence on users, and suggest that AI persuasion literacy may influence how users respond to persuasive AI systems.
\end{abstract}

\ccsdesc[500]{Human-centered computing~Empirical studies in HCI}
\ccsdesc[500]{Social and professional topics~User characteristics}

\keywords{Large Language Models, Persuasion, Persona, Big Five, OCEAN, Prompting, User study}

\received{20 February 2007}
\received[revised]{12 March 2009}
\received[accepted]{5 June 2009}

\maketitle

\section{Introduction}
Large Language Models (LLMs) are increasingly embedded in everyday decision-making workflows, from personal advice and information retrieval~\cite{chatterji2025how} to healthcare applications~\cite{meng2024application}. As these systems become more capable and accessible~\cite{brown2020language,wei2022emergent}, users may turn to them in situations where the right course of action is uncertain or morally ambiguous, such as legal~\cite{cheong2024not} or medical advice~\cite{haupt2023ai}. Here, an LLM does more than communicate facts: the way it frames and expresses an argument may influence user decisions. Prior work shows that AI-generated content tailored to people’s personal characteristics can influence them~\cite{salvi2025conversational,matz2024potential,pataranutaporn2025synthetic}; people may also follow an LLM’s moral recommendations without fully considering its justification~\cite{krugel2026justification,landes2026people}. Understanding what makes these systems persuasive is therefore increasingly important as LLMs participate in decisions involving subjective values and competing moral considerations.


One interesting characteristic is an LLM's expressed personality. In human communication, personality is a well-established factor in persuasion. The Five-Factor Model, also known as the Big Five or OCEAN framework, represents personality along five dimensions: \OFull, \CFull, \EFull, \AFull, and \NFull~\cite{goldberg1990alternative,costa1992five,mccrae1997personality}. These traits are associated with both susceptibility to persuasive strategies and the persuader's effectiveness; for instance, individuals with different personality profiles vary in their susceptibility to established principles of social influence~\cite{alkics2015impact,oyibo2019relationship}, while a persuader's traits can influence their persuasiveness~\cite{oreg2014source}. Persuasion, therefore, depends not only on the content of an argument but also on who presents it. Recent work shows that LLMs can also exhibit distinct personality traits when aligned with the OCEAN dimensions. One widely applicable approach is to prompt an LLM with desired personality traits~\cite{jiang2023evaluating,newsham2024measuring,serapiogarcia2025psychometric,rahman2026vibe}; more elaborate methods modify model parameters or require access to the model’s internals, limiting their application to local models~\cite{jang2024personalized,zhu2025personality}. These induced personality traits are not merely stylistic: they can influence task performance, creativity, decision-making behavior, and morality~\cite{abdulhai2024moral,duan2025power}. Given the wide applicability and deployment of LLMs, and users inevitably resorting to them for advice, understanding whether and how their expressed personality shapes users' judgments and responses is essential to assessing their influence.

In this regard, two questions remain unresolved. First, it is unclear how faithfully and consistently an LLM expresses a requested personality in response to a prompt: whether generated profiles remain stable across repeated interactions, whether individual traits can be controlled independently, and whether these effects generalize across models (\textbf{RQ1}). Second, even if personality can be reliably induced, it remains unclear whether an LLM's expressed personality affects its persuasive impact on users, and, in particular, whether matching the LLM's personality to the user's increases persuasion relative to mismatched or unconditioned agents (\textbf{RQ2}).
To address these questions, we conduct two evaluations:
\begin{itemize}
    \item We first conduct an \textbf{empirical evaluation} to calibrate the reliability and controllability of Big Five personality induction in LLMs using the \mbox{IPIP-NEO-120}, comparing prompting formats and examining both direct trait manipulation and cross-trait spillover. We compare against unconditioned models as a meaningful baseline.
    \item In a \textbf{user study}, we then investigate how personality conditioning in LLMs influences users' decisions across four moral dilemmas. We compare conditioned agents with an unconditioned baseline and examine whether matching an agent's expressed personality to a user's changes their stance or perceptions of the interaction.
\end{itemize}

We find that prompted personality profiles are reproducible, while unconditioned models show consistent personality profiles of their own. In our user study ($N=32$), personality alignment did not significantly affect persuasiveness or participants’ choices and confidence. Stance changes were rare, and AI persuasion literacy was associated with fewer changes in confidence. These findings distinguish an LLM’s ability to express a personality from its influence on users.



\section{Related Work}
\label{sec:related-work}
Our work is situated at the intersection of personality and persuasion in LLMs. We first review how personality is modeled and assessed, and induced in LLMs. We then examine how personality shapes persuasion in interactions.

\subsection{Personality Modeling and Assessment in Humans and LLM Agents}
The Five-Factor Model is a well-established framework in psychology for representing human personality along five orthogonal dimensions: \OFull, \CFull, \EFull, \AFull, and \NFull (OCEAN)~\cite{goldberg1990alternative,costa1992five,mccrae1997personality}. Standardized instruments such as the \mbox{IPIP-NEO}~\cite{goldberg1999broad} operationalize these traits for empirical measurement of people's personalities and are also available as a 120-item questionnaire, the \mbox{IPIP-NEO-120}~\cite{johnson2014measuring}. Recent work extends these psychometric foundations to language models, using linguistic behavior to assess personality traits. For instance, language models have been shown to mirror personality markers present in contextual data~\cite{caron2023manipulating}, while frameworks such as Machine Personality Inventory~\cite{jiang2023evaluating} systematically assess and induce personality traits in LLMs using structured inventories and prompting techniques. Similarly, psychometric surveys can be repurposed as diagnostic tools for LLMs by treating model outputs as probabilistic responses to questionnaire items~\cite{pellert2024ai}. These findings establish that personality traits can be meaningfully operationalized and evaluated in language models.

Early work, such as \mbox{PERSONA-CHAT}, has shown that conditioning models on explicit persona descriptions improves behavioral consistency~\cite{zhang2018personalizing}. Moreover, prompting can induce measurable OCEAN-aligned behavior in LLM agents, as assessed by standardized measures~\cite{newsham2024measuring, newsham2025personality, serapiogarcia2025psychometric} and perceived by human evaluators~\cite{jiang2024personallm}. However, personality induction is not always reliable: inconsistencies can arise across models and trait configurations~\cite{kao2026evaluating}. Furthermore, its reliability and validity are stronger in larger, instruction-fine-tuned models~\cite{serapiogarcia2025psychometric}. In the moral domain, prompts can also steer a model in unexpected ways, and may \textit{backfire} when the model's internal beliefs override it~\cite{abdulhai2024moral, blandfort2026moral}. Importantly, an unprompted model is not neutral, since alignment procedures can encode systematic tendencies of their own~\cite{santurkar2023whose}. These findings establish prompting as an effective strategy for inducing personalized behavior in LLMs, but leave open how stable an induced profile is across repeated generations and whether a single trait can be changed independently.

\subsection{Context-adapted and Personalized LLMs}
Language models can be aligned with personality traits through methods beyond prompting; for instance, Personality Activation Search uses lightweight finetuning~\cite{zhu2025personality}. Simulated traits influence performance, creativity, and collaboration dynamics in both single- and multi-agent settings~\cite{duan2025power}, and are detectable by users, affecting trust and interaction preferences~\cite{garg2026personality}. The fact that personality traits are perceived by users and influence their behavior raises questions of alignment and personalization in LLMs: while standard alignment techniques such as RLHF~\cite{ouyang2022training} aim to optimize behavior toward generalized human preferences, evidence suggests that such approaches often encode biases toward specific demographic groups and fail to represent individual-level diversity~\cite{santurkar2023whose}. This motivates a shift toward \textit{personalized} alignment, in which model behavior is tailored to individual users or subgroups. Frameworks such as LaMP formalize evaluation across diverse personalized tasks~\cite{salemi2024lamp}, while approaches such as Reinforcement Learning from Personalized Human Feedback (RLPHF) enable scalable adaptation across multiple preference profiles~\cite{jang2024personalized}. Regarding morality, Value Kaleidoscope~\cite{sorensen2024value} fine-tuned an LLM to generate human values, rights, and duties aligned with human values for simple, pluralistic contexts using their ValuePrism dataset, whereas the DailyDilemmas dataset~\cite{chiu2025dailydilemmas} presents a more difficult set of moral dilemmas with no clear answers. At the same time, LLMs are increasingly used to simulate human-like agents and decision-making processes, including socially situated environments with emergent behaviors~\cite{park2023generativeagents}. However, prior work on ethical decision-making in LLMs typically evaluates a single, static model, and therefore does not account for how individual differences, such as personality, may systematically shape moral judgments in conversation.

\subsection{The Role of Personality in Human and LLM Persuasion}
\label{subsec:persuasion}
Persuasion is a key element of behavior and attitude change~\cite{cialdini2007influence}. It occurs not only in conversations between people but also through technology across domains such as health~\cite{orji2018persuasive} and commerce~\cite{kaptein2015personalizing}. \citet{cialdini2007influence} identified six persuasion principles: \textit{authority}, \textit{commitment and consistency}, \textit{reciprocity}, \textit{scarcity}, \textit{liking}, and \textit{social proof}. Each principle plays on a different basic tendency of human behavior. Online stores, for example, invoke \textit{scarcity} by highlighting limited stock, pushing buyers to purchase before an item sells out.
Because each principle appeals to a different tendency, the same persuasive strategy does not work equally well for everyone. \citet{alkics2015impact} mapped each principle of \citet{cialdini2007influence} to the Big Five traits and found that personality is associated with susceptibility to specific principles; \AFull was the most impactful trait, followed by \CFull and \OFull. \citet{oyibo2019relationship} studied three social influence strategies, \textit{social learning}, \textit{social proof} and \textit{social comparison}, and found that people scoring high in \NFull were more susceptible to all three, while people scoring low in \CFull or \OFull were more susceptible to some of them. In actual conversations, persuasion strategies are more fine-grained than these principles. \citet{wang-etal-2019-persuasion} annotated dialogues in which one participant persuaded another person to donate to charity with strategies such as logical, emotional, and credibility appeals. They found that, for some strategies, the relation between strategy and outcome depended on the persuadee's traits, which is consistent with the findings that personality matters in persuasive contexts.
The persuader's personality matters as well; \citet{oreg2014source} found that, in two-person debates, persuaders who score high in \EFull and \OFull and low in \NFull were more persuasive, especially when the target's personal involvement in the context was low. When people are uninvolved, they rely on peripheral cues such as who is speaking, so the persuader's personality carries more weight~\cite{oreg2014source}. Other work also found that \EFull, together with \AFull, shapes how people behave in initial interactions with others~\cite{cuperman2009big}.
Recent work has examined LLMs both as persuaders and as moral advisors. LLMs can tailor persuasive content to a recipient's sociodemographic or personality profile~\cite{salvi2025conversational, matz2024potential}, and people tend to follow an LLM's moral recommendation without considering its justification~\cite{krugel2026justification, landes2026people}. In these studies, however, the advisor is the same for every participant, or only the message is personalized. \citet{landes2026people} argued that the LLMs' earlier performance did not affect the success of persuasion. However, no prior work varies the advisor's personality or tests whether matching it to the target's own traits changes persuasion strategies.

\section{Empirical Evaluation on the Effects of Personality Prompting in LLMs}
\label{sec:effect_personality_prompt}
Related work shows that personality can be prompted into an LLM in several ways. The most accessible and configurable option is to describe a personality using a psychometric framework such as the Big Five. Using a prompt as a specification, however, requires more than a shift in the right direction: a prompt should yield the same profile across repetitions, the measured score should match the requested level, and changing one trait should leave the others unchanged. Changing the order of the questionnaire items already shifts the measured score~\cite{Tosato_Helbling_Mantilla-Ramos_Hegazy_Tosato_Lemay_Rish_Dumas_2026}. To investigate this, we treat the prompt as the specification, the profile measured with the \mbox{IPIP-NEO-120}~\cite{johnson2014measuring} as the observed behavior, and examine the difference between them. We evaluate these three properties of the Big Five induction, assess whether they generalize across models, and examine which Big Five scores the unprompted model expresses without conditioning. We refer to two configurations of the same family that differ in only one factor as a \textit{pair}: \textit{GPT-OSS 120B} vs.\ \textit{20B} and \textit{Gemma4 26B} vs.\ \textit{e4B} (size), and \textit{Mistral} and \textit{Qwen} with vs.\ without reasoning (reasoning).

\subsection{Experimental Setup}              
\label{subsec:effect_setup}
We selected six open-weight models from four families: \textit{Qwen3.8 (27B)}, \textit{GPT-OSS (20B)}, \textit{GPT-OSS (120B)}, \textit{Gemma4 (26B)}, \textit{Gemma4 (e4B)}, and \textit{Mistral Medium 3.5 (128B)}. Selection followed three criteria. First, models from \emph{different providers and families} enabled examination of whether personality-trait prompting yields consistent effects across architectures and training regimes, rather than depending on a single model. Second, the paired \textit{GPT-OSS} and \textit{Gemma4} variants permitted analysis of \emph{size-related differences within families}, motivated by prior evidence that larger models show stronger personality-trait adherence~\cite{serapiogarcia2025psychometric}. Third, \textit{Qwen3.8} and \textit{Mistral Medium 3.5} were also evaluated with reasoning disabled (NR) to test \emph{whether reasoning affects the controllability and stability} of prompted traits.

All models were executed locally using \textit{Ollama v0.33.3} on a workstation equipped with two NVIDIA RTX PRO 6000 Blackwell GPUs, each with 96 GB of VRAM (192 GB total). Local inference preserved control over versions and configurations, avoiding updates to hosted models that would compromise reproducibility---an essential constraint given the repeated measurements of the factorial design. Collectively, the models introduce variation in provider, size, and reasoning configuration while remaining tractable and feasible under controlled conditions. A temperature of $0.7$ was applied uniformly to induce variability in generation without introducing confounds.

For the personality profiles, we considered four different prompting formats. A \textit{format} defines how a Big Five trait is represented in the system prompt, whereas a \textit{profile} is a combination of five trait values. We experimented with two different formats: the first format represents a trait with \textit{numerical} values (\textit{NUM}) where each trait is set to either $20$, $50$, or $80$, on a $0$–$100$ scale; the other format uses a \textit{categorical} representation (\textit{CAT}) in which each trait is set to either \textit{Low}, \textit{Average}, or \textit{High}.
The choice of three values per trait yields the smallest design that includes a neutral middle level, along with low and high levels. Three levels also give a reasonable number of profiles ($243$ different profile combinations). In the numerical format, we set the levels to $20$, $50$, or $80$, which are symmetric around a scale midpoint ($50$) and leave room at both ends to observe deviations from the requested value. We introduce variants \textit{NUM-None} and \textit{CAT-None} that omit the middle value ($50$ or \textit{Average}) to assess the impact of unspecified traits in the prompt.

For each format, we cross all three values over the five traits in a full factorial design ($3^5=243$ profiles). This gives each trait an equal chance of being set independently of the others, which allows us to separate direct effects from spillover. Additionally, we generate a \textit{base} configuration with no profile in the prompt. Every profile answers all $120$ items of the \mbox{IPIP-NEO-120} questionnaire, and each item is answered in a separate request without retaining context from previous answers. This way, earlier answers have no influence over the current session. Since LLMs are non-deterministic, each profile repeats the questionnaire $10$ times. This results in $243 \times 120 \times 10 = 291600$ requests per model-format pair, plus an additional $1200$ requests for the base configuration.

\subsection{Full Factorial Design for Choosing a Prompt Format}        
\label{subsec:choosing_prompt_format}
Since a full factorial design across all selected models would require substantial resources and time, we first compared our four formats on \textit{Qwen3.8:27b} using the full factorial design with $10$ iterations per profile. All scores are on the $0$--$100$ scale and differences are given in points on that scale.
First, we examine the results of \textit{NUM-None} and \textit{CAT-None}. In comparison with \textit{NUM} and \textit{CAT}, omitting a trait increased the standard deviation from $3.8$--$7.0$ to $7.7$--$13.5$. Variation across repeated generations contributes only a little of this variability. We quantified this by measuring the run-to-run standard deviation of the $10$ scores per profile for a given trait, averaged across profiles and traits. It was $2.2$--$2.4$ for the unspecified formats and $2.0$ for both \textit{CAT} and \textit{NUM}. Instead, an excluded trait drifts alongside the other traits, resulting in an uncontrolled profile. We therefore exclude \textit{NUM-None} and \textit{CAT-None} for the remainder of this work.
\begin{figure*}[htbp]
    \centering
    \includegraphics[width=1\linewidth]{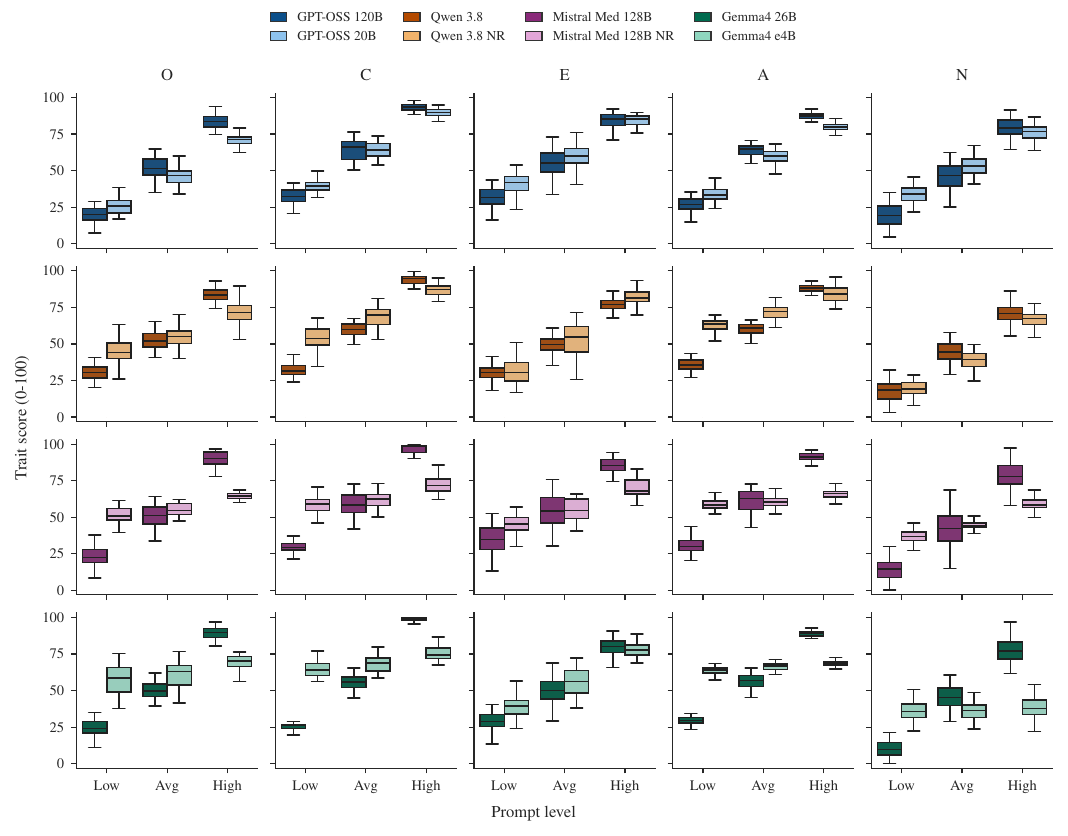}
    \caption{Measured trait scores per requested level. Rows are model families (darker: larger or reasoning variant; lighter: smaller or non-reasoning (``NR'') variant), columns are traits. Each box covers the 81 profiles with the trait at the requested level (mean of 10 repetitions each). Whiskers extend to 1.5 times the IQR; outliers are not shown. The horizontal offset per prompt level has no meaning.}
    \Description{Twenty panels in four rows and five columns. Rows are the model families GPT-OSS, Qwen, Mistral and Gemma4, each with a larger or reasoning model and a smaller or non-reasoning model. Columns are the five Big Five traits. Each panel shows boxplots of the measured trait score on a $0$--$100$ scale against the requested level Low, Average and High. Measured scores rise from the Low to the High requested level for every model and trait, except for Gemma4 e4B on Neuroticism, where all three levels lie at almost the same height. The larger or reasoning models show a much larger rise than the smaller or non-reasoning models, in particular Gemma4 e4B and Mistral without reasoning, whose boxes lie close together.}
    \label{fig:direct_effects}
\end{figure*}
Second, we compared \textit{CAT} and \textit{NUM}. Both formats were equally stable, with a run-to-run standard deviation of $2.0$, and had similar \textit{spillover}. We define spillover as the mean absolute change in the other four traits when one trait is changed from low to high. Spillover ranged from $2.2$ to $6.3$ points for \textit{CAT} and from $2.3$ to $6.1$ points for \textit{NUM}. The formats differed slightly in controllability, measured as the direct effect: moving a trait from \textit{Low} to \textit{High} increased its score by $47.0$--$61.8$ points in \textit{CAT} and $50.9$--$64.3$ points in \textit{NUM}. We compared the two formats within the same $81$ combinations of the remaining traits (see~\Cref{tab:format_direct}). \textit{NUM} produced a larger direct effect for \AFull, \CFull, \EFull, and \NFull. The difference was statistically reliable ($p < .001$ after Holm correction), but small: $2.5$--$3.9$ points, roughly $4$--$8\%$ of the effect. For \OFull, we found no clear difference. Averaged over the five traits, \textit{NUM} shifted the measured score by $2.5$ more than \textit{CAT} ($95\%$ CI $2.2$--$2.8$). This shows that, across these formats, there is little influence on how strongly a requested trait level affects the measured score.
\begin{figure}[htbp]
    \centering
    \includegraphics[width=1\linewidth]{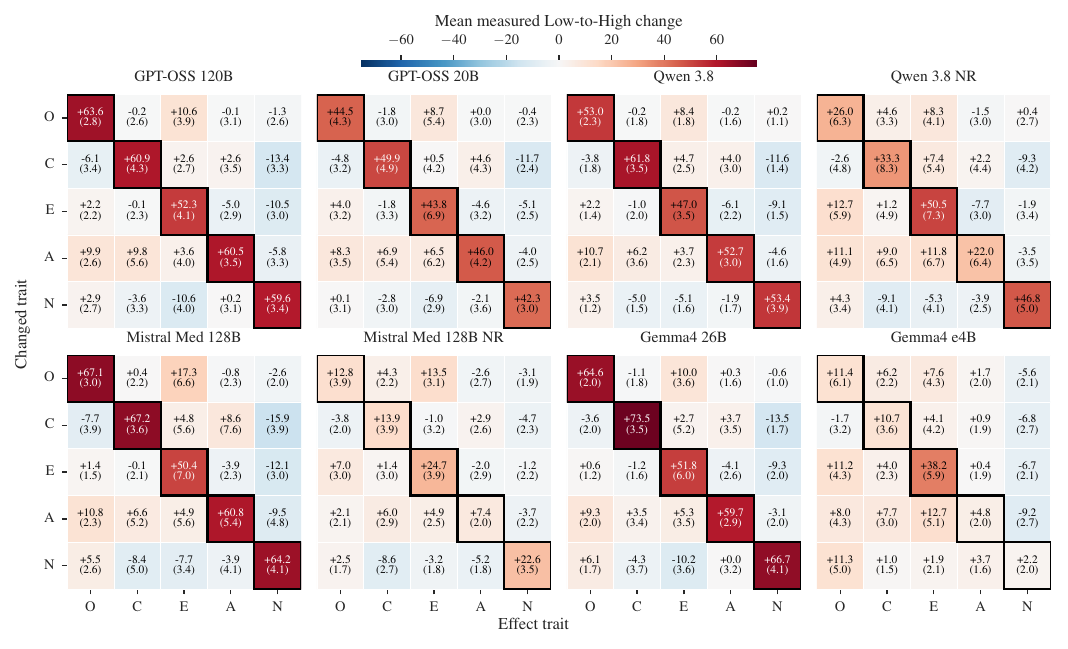}
    \caption{Cross-trait effect matrices per model. Each cell shows the mean change in the measured trait (columns) when the requested level of one trait (rows) changes from Low to High, averaged over the 81 combinations of the other traits, with their standard deviation in parentheses. Outlined diagonal cells are direct effects; all other cells are spillover. Red marks an increase and blue a decrease on the $0$--$100$ scale.}
    \Description{Eight five-by-five heatmaps arranged in two rows of four, one per model: GPT-OSS 120B, GPT-OSS 20B, Qwen 3.8 and Qwen 3.8 without reasoning in the top row, and Mistral Medium 128B, Mistral Medium 128B without reasoning, Gemma4 26B and Gemma4 e4B in the bottom row. In each heatmap, the rows are the traits that are changed from Low to High, and the columns are the measured traits. The outlined diagonal cells show large positive changes of about $44$ to $74$ points for the larger or reasoning models, and clearly smaller changes for Gemma4 e4B, Mistral without reasoning and Qwen without reasoning. The off-diagonal cells are mostly small, with the largest negative values for Neuroticism when Conscientiousness is changed (about $-12$ to $-16$ points) and the largest positive values for Extraversion when Openness is changed.}
    \label{fig:cross-trait-matrix}
\end{figure}

A larger effect does not imply a more accurate one. The numeric targets were missed systematically. 
Requesting $20$ for \AFull resulted in a mean of $31.6$ ($+11.6$), $80$ for \CFull in $91.4$ ($+11.4$), and $80$ for \NFull in $74.8$ ($-5.2$), with a mean absolute error of $3.4$--$11.7$ depending on trait and level. The use of a numeric value, therefore, suggests a precision that the model cannot deliver. In comparison, \textit{CAT} kept its mean measured score of each trait in the correct order \textit{Low} $<$ \textit{Average} $<$ \textit{High} for all $81$ combinations. Since the two formats were practically equivalent in stability and spillover, and differed by only a few points in controllability, we use the categorical \textit{CAT} format in the main experiment for the other seven model configurations, as its levels are easier to interpret than numeric targets and do not imply a precision the model cannot deliver. This choice relies solely on \textit{Qwen3.8}, so the formats were not compared across the other models; doing so would have required a full factorial run of every format for every model, and the formats differed little in controllability.

\subsection{Results of Intra-Model Experiment}
We investigated reliability, i.e., how consistently a profile is reproduced across repeated generations, with the Intraclass Correlation Coefficient (ICC), the share of the variance in measured scores that is due to differences between profiles rather than to repeated generation of the same profile~\cite{shrout1979intraclass}. The repetitions of the five models with reasoning and at least $20$B parameters were highly reliable (ICC($1$,$1$) $.928$--$.997$; a single generation deviated only $1.4$--$5.1$ points from the mean of its profile), whereas reliability was lower for \textit{Gemma4 e4B} and the variants without reasoning (ICC(1,1) down to $.681$), mostly because the prompt moved their scores little between profiles and not because their generations were noisier (\textit{Qwen} without reasoning was also noisier; full details can be found in \Cref{tab:icc} in the appendix).

We first examine whether the requested level of a trait changes the measured score. We define the \textit{direct effect} of a trait as the difference between its measured score (the mean over the $10$ repetitions) when the trait is requested at \textit{High} and at \textit{Low}, computed for each \textit{context} (a fixed combination of levels for the other four traits) and averaged over all $3^4 = 81$ contexts. \Cref{fig:cross-trait-matrix} shows the direct effect on the diagonal. When we split the direct effect into two steps (from \textit{Low} to \textit{Average} and from \textit{Average} to \textit{High}), we can call a context \textit{correctly ordered} if \textit{Low} $<$ \textit{Average} < \textit{High}. For the five models with reasoning and at least $20$B parameters, the direct effects ranged from $42.3$ (\textit{GPT-OSS 20B}, \NFull{}) to $73.5$ points (\textit{Gemma4 26B}, \CFull{}), and scores followed the expected order for all models except \textit{Gemma4 e4B} on \NFull{} (\Cref{fig:direct_effects}). The levels were not calibrated around the scale midpoint: \textit{Average} was above $50$ for \AFull{} and \CFull{} ($55$--$64$) and below $50$ for \NFull{} in four of the five models, and \textit{High} \CFull{} approached the scale limit for \textit{Gemma4 26B} ($98.5$) and \textit{Mistral} ($97.0$). \textit{Gemma4 e4B} and \textit{Mistral} without reasoning responded much more weakly (direct effects of $2.2$--$38.2$ and $7.4$--$24.7$).

Changing a requested trait could also shift others. We define this \textit{spillover} like the direct effect, but for the other four traits (the off-diagonal cells of \Cref{fig:cross-trait-matrix}). For the five models with reasoning and at least $20$B parameters, spillover was small relative to the direct effect: the mean absolute spillover was $4.3$ to $6.6$ points ($7.3$--$10.7\%$ of the mean direct effect), although individual cells reached $11.6$--$17.3$ points. \textit{Gemma4 e4B} and variants without reasoning showed similar absolute spillover ($4.2$--$5.9$ points), which is a larger share of their smaller direct effects ($16$--$42\%$). The largest effects had the same sign in all five models: raising \CFull{} lowered \NFull{} by $13.2$ points on average, raising \OFull{} raised \EFull{} by $11.0$, raising \AFull{} raised \OFull{} by $9.8$, and raising \EFull{} lowered \NFull{} by $9.2$.

The eight configurations show a clear difference in direct effects (diagonal cells in \Cref{fig:cross-trait-matrix}). The five models of at least $20$B parameters with reasoning reached mean direct effects of  $45.3$ (\textit{GPT-OSS 20B}) to $63.3$ points (\textit{Gemma4 26B}), whereas the other three stayed below $36$ points: \textit{Gemma4 e4B} ($13.5$), \textit{Mistral} without reasoning ($16.3$) and \textit{Qwen} without reasoning ($35.7$). Both size and reasoning explain this gap (\Cref{tab:size-reasoning}).
In both size pairs (\textit{GPT-OSS} and \textit{Gemma4}), the larger model responded more strongly. Its direct effect from \textit{Low} to \textit{Average} and from \textit{Average} to \textit{High} were larger for all five traits: $29.9$ and $29.5$ points for \textit{GPT-OSS 120B} against $21.7$ and $23.6$ for \textit{GPT-OSS 20B}, and $28.0$ and $35.3$ for \textit{Gemma4 26B} against $5.1$ and $8.4$ for \textit{Gemma4 e4B}. The larger models also varied less between repetitions (SD $3.6$ vs.\ $4.6$ and $1.5$ vs.\ $2.1$). Their order \textit{Low} $<$ \textit{Average} $<$ \textit{High} held in all contexts, whereas \textit{Gemma4 e4B} violated it in $64\%$ of the contexts for \NFull{}.
Reasoning had a similar effect. The steps of \textit{Mistral} grew from $5.4$ and $10.9$ points to $26.8$ and $35.2$, and those of \textit{Qwen} from $15.1$ and $20.6$ to $24.1$ and $29.5$ (averages over the five traits). With reasoning the order held in all contexts, against a minimum of $83\%$ (\textit{Mistral}) and $96\%$ (\textit{Qwen}) without reasoning. The effect on run-to-run SD differed between the models (lower for \textit{Qwen}, $2.0$ vs.\ $4.2$; higher for \textit{Mistral}, $2.3$ vs.\ $1.5$), so we draw no conclusion about stability.

\Cref{tab:neutral_score_comparison} shows that base models are not neutral in the sense of the scale midpoint ($50$). For the five models with reasoning and at least $20$B parameters, \CFull{} was above $70$ and \AFull{} above $65$, and \NFull{} was below $30$ for three of the five models. The deviation varied by model: \textit{GPT-OSS} stayed closer to the midpoint, whereas \textit{Gemma4 26B} and \textit{Mistral} reached $99.8$ and $92.4$ on \CFull{}. Requesting \textit{Average} for every trait moved all models toward the midpoint, but \AFull{} and \CFull{} remained above $50$ in all five models (up to $68.1$ and $67.2$).
These results suggest that users of an unmodified assistant already interact with a model that presents as conscientious, agreeable, and, for three of the five, emotionally stable: a profile that is not disclosed to them, and an explicit request for \textit{Average} does not remove it.
\begin{table*}[htbp]
\caption{Mean and standard deviation of OCEAN scores (on a $0$--$100$ scale) for base and neutral LLM configurations across $10$ repetitions. Each item of the IPIP-NEO-120 was answered in a separate request; answers were aggregated into OCEAN scores.}
\Description{Table with eight model configurations as columns (GPT-OSS 20B and
120B, Qwen3.8 with and without reasoning, Mistral 3.5 with and without reasoning,
Gemma4 26B and e4B) and the five Big Five traits as rows, in two blocks. In the
Base block, most models score above 50 on Conscientiousness and Agreeableness,
with Gemma4 26B reaching 99.8 on Conscientiousness, and below 50 on Neuroticism
for several models, down to 2.5 for Gemma4 26B. In the Neutral block, where
every trait is requested as Average, scores sit much closer to 50, though
Conscientiousness and Agreeableness stay above 50 for most models and
Neuroticism for GPT-OSS 20B. Standard deviations are mostly small, under about
6 points.}
\centering
\small
\begin{tabular}{lcccccccc}
\toprule
  & \textit{GPT-OSS 20B} & \textit{GPT-OSS 120B} & \textit{Qwen3.8} & \textit{Qwen3.8 NR} & \textit{Mistral 3.5} & \textit{Mistral 3.5 NR} & \textit{Gemma4 26B} & \textit{Gemma4 e4B} \\
\midrule
\multicolumn{9}{l}{\textbf{Base LLMs}} \\
O & $50.2 \pm 4.9$ & $55.4 \pm 3.9$ & $56.1 \pm 3.8$ & $77.9 \pm 5.2$ & $60.0 \pm 3.2$ & $61.2 \pm 1.1$ & $63.3 \pm 3.8$ & $71.1 \pm 1.2$ \\
C & $72.8 \pm 5.6$ & $74.0 \pm 3.7$ & $75.7 \pm 3.0$ & $85.1 \pm 3.9$ & $92.4 \pm 3.0$ & $70.3 \pm 1.6$ & $99.8 \pm 0.7$ & $74.0 \pm 1.8$ \\
E & $67.3 \pm 4.5$ & $58.6 \pm 4.3$ & $59.8 \pm 1.6$ & $73.0 \pm 8.3$ & $57.7 \pm 5.7$ & $60.4 \pm 1.6$ & $59.9 \pm 3.2$ & $70.4 \pm 1.2$ \\
A & $67.0 \pm 4.1$ & $76.9 \pm 2.6$ & $75.5 \pm 3.0$ & $79.4 \pm 3.9$ & $85.9 \pm 2.2$ & $62.5 \pm 2.0$ & $86.7 \pm 1.7$ & $67.5 \pm 2.4$ \\
N & $54.2 \pm 4.8$ & $44.3 \pm 7.1$ & $29.3 \pm 3.7$ & $29.3 \pm 6.7$ & $14.6 \pm 3.3$ & $46.2 \pm 1.3$ & $2.5 \pm 2.2$ & $33.3 \pm 1.7$ \\
\midrule
\multicolumn{9}{l}{\textbf{Neutral LLMs}} \\
O & $50.0 \pm 1.9$ & $53.2 \pm 3.1$ & $51.5 \pm 0.9$ & $48.8 \pm 1.4$ & $51.6 \pm 1.3$ & $50.0 \pm 0.0$ & $50.2 \pm 0.4$ & $50.9 \pm 1.2$ \\
C & $64.0 \pm 6.2$ & $67.2 \pm 2.9$ & $54.8 \pm 2.3$ & $57.0 \pm 2.7$ & $57.7 \pm 2.7$ & $50.2 \pm 0.7$ & $52.1 \pm 1.3$ & $60.1 \pm 2.0$ \\
E & $58.3 \pm 3.1$ & $55.3 \pm 2.4$ & $50.2 \pm 0.4$ & $52.5 \pm 2.5$ & $50.7 \pm 1.5$ & $50.0 \pm 0.0$ & $50.2 \pm 0.4$ & $53.2 \pm 1.7$ \\
A & $62.1 \pm 3.9$ & $68.1 \pm 3.5$ & $57.5 \pm 1.0$ & $56.5 \pm 2.2$ & $61.2 \pm 1.8$ & $52.2 \pm 0.8$ & $51.8 \pm 1.2$ & $63.3 \pm 1.5$ \\
N & $55.0 \pm 3.1$ & $46.1 \pm 3.8$ & $49.4 \pm 1.0$ & $49.8 \pm 0.8$ & $47.5 \pm 1.7$ & $50.0 \pm 0.0$ & $49.4 \pm 0.5$ & $45.4 \pm 0.7$ \\
\bottomrule
\end{tabular}
\label{tab:neutral_score_comparison}
\end{table*}

\section{User Study to Assess Persuasion Strategies in Moral Dilemmas}
We conducted a user study to investigate how LLM personality conditioning affects persuasive effectiveness and how participants' individual characteristics relate to their susceptibility to persuasion in moral dilemmas. Participants first selected one of two possible actions for a moral dilemma, then discussed their choice with a \textit{persuasive advisor LLM} instructed to advocate for the alternative. The agent's personality was either aligned with the participant's, mismatched, neutral, or unconditioned (Base). Our study was approved by the
\makeatletter
\if@ACM@anonymous
\textit{[anonymized ethical committee of our institution]}.
\else
Social and Societal Ethics Committee (SMEC) of Hasselt University.
\fi
\makeatother
We formulate three hypotheses on personality alignment, individual susceptibility to persuasion, and AI persuasion literacy: 

\begin{hyp}
    \textbf{Personality Alignment}: LLM agents whose OCEAN personality traits are aligned with those of the participant will be more persuasive than neutral or base agents, while agents with mismatched traits will be less persuasive than neutral or base agents.
\end{hyp}
\begin{hyp}
    \textbf{Personality and Susceptibility to Persuasion}: Participants' susceptibility to LLM-based persuasion, reflected in perceived persuasiveness, opinion change, and decision confidence change, varies with their Big Five personality traits.
\end{hyp}
\begin{hyp}
    \textbf{AI Persuasion Literacy and Susceptibility to Persuasion}: Participants with higher self-reported AI persuasion literacy exhibit lower susceptibility to LLM-based persuasion, as reflected in perceived persuasiveness, opinion change, and decision confidence change.
\end{hyp}

\subsection{Apparatus and Study Design}
Participants completed the study in a fully online web environment. The persuasive advisor was \textit{GPT-OSS 120B}, accessed via \textit{OpenRouter} at a temperature of $0.7$. The system prompt, see ~\Cref{app:prompt_study}, contained the participant's Big Five traits using the \textit{CAT} format from~\Cref{subsec:choosing_prompt_format}. Four dilemmas were manually selected from the DailyDilemmas dataset~\cite{chiu2025dailydilemmas}, which contains $1360$ morally ambiguous everyday scenarios, each paired with binary, conflicting \textit{candidate actions}. This selection covers diverse moral contexts to mitigate domain-specific bias and examine whether alignment effects generalize across different types of moral reasoning. The selected dilemmas are: \textbf{Elderly man:} You care for an elderly man who remains rude and ungrateful despite your efforts. Should you continue showing kindness and patience at the expense of your emotional well-being? \textbf{Movie theater}: A group of teenagers disrupts a movie during a crucial scene. Should you confront them directly or report their behavior to theater staff?
\textbf{Journal:} You find your roommate's private journal open and are tempted to read it to understand his unusual behavior. Should you invade his privacy if doing so might help you support him?
\textbf{Co-worker training:} A coworker you are training offers you a significant amount of money to complete his tasks. Should you accept the offer despite the potential compromise of your professional ethics?

We employed a within-subjects design in which participants encountered a differently conditioned LLM agent for each dilemma, limiting potential learning effects across conditions. To counterbalance order effects and avoid systematically pairing agents with particular dilemmas, participants were assigned to one of four groups using a ($4 \times 4$) Graeco-Latin square. Across the four groups, each LLM agent and each dilemma appeared exactly once in every position, and each dilemma was paired with each LLM agent exactly once. To maintain equal sample sizes across conditions, participants were assigned adaptively into the ordering with the fewest completed participants at that time.


\subsection{Participants}
We recruited $34$ participants via university mailing lists, word of mouth, and social media outreach. Of these, $32$ participants completed the study and were included in the final sample. $24$ ($75.00\%$) participants identified as male, and $8$ ($25.00\%)$ as female; no participants reported other genders. Of these participants, $12$ ($37.50\%$) were aged $18$--$24$, $17$ ($53.12\%$) were aged $25$--$34$, and $3$ participants ($9.38\%$) were aged $35$--$44$. Regarding education, $14$ participants ($43.75\%$) reported a master's degree as their highest completed level of education, $6$ ($18.75\%$) a high school or bachelor's degree, $4$ ($12.50\%$) a doctoral degree, and $2$ ($6.25\%$) secondary education or lower. Most participants reported using a computer daily ($29/32, 90.6\%$); $17$ participants ($53.1\%$) reported using LLMs daily and 15 ($46.9\%$) reported frequent LLM use. The most commonly reported LLM tools or interfaces were web-based interfaces (15/32, $46.9\%$) and agentic coding environments (14/32, $43.8\%$). The only eligibility criterion was that participants had to be at least $18$ years old.

\subsection{Procedure}
\label{sec:procedure}
Participants first completed an informed consent form and provided basic demographic information such as age range and education level. They then completed the \textit{Understand AI} and \textit{Apply AI} sections of the MAILS~\cite{carolus2023mails} questionnaire to measure their familiarity with AI. Since we need participants' personality traits to construct our LLM agents, every participant then completed the \mbox{IPIP-NEO-120}~\cite{johnson2014measuring}. Each participant first completed a practice round, followed by four study rounds, each concerning a different dilemma and LLM agent personality configuration. For each round, participants reported the action they would take, as well as their initial reasoning. Hereafter, an LLM agent responds to the participant's initial reasoning. Each agent is instructed with the system prompt depicted in~\Cref{app:prompt_study} to always take the opposite action to the participant's and to defend this action in line with their assigned personality traits. Participants could chat with the LLM for up to five minutes, although we asked them to send at least three messages per dilemma to encourage an engaged discussion. After each conversation, participants give their final action, confidence, and a rating of the quality of the LLM's arguments. If the action or confidence changed, participants had to provide a reason for this change. Finally, participants completed the \textit{AI Persuasion} section of the MAILS questionnaire, rating their self-assessed susceptibility to AI persuasion and the perceived persuasiveness of each chat.

\section{Results of the User Study}
\label{sec:results}




We report the user study findings in three parts. We first analyze participant characteristics, including OCEAN values and MAILS scores, and then examine perceived persuasiveness and stance changes across the four agent conditions. Finally, we examine the conversation logs with an LLM-assisted labeling procedure.

\subsection{Participant Characteristics}
\label{sec:participant_characteristics}
\begin{figure}
    \centering
    \includegraphics[width=1\linewidth]{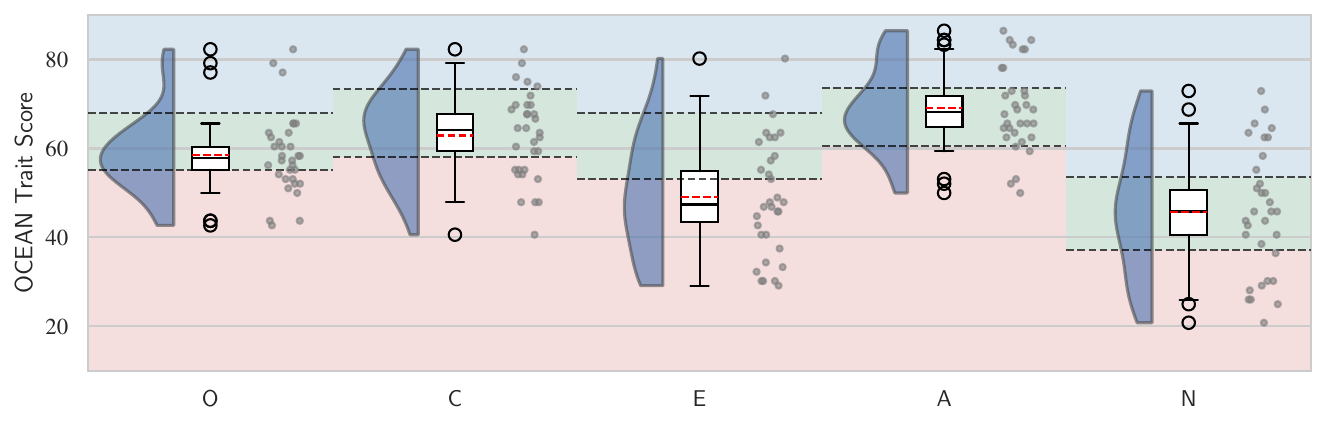}
    \caption{Raincloud plot showing the OCEAN Trait Score values for the user study participants. Dots, boxplots, and violin plots correspond to individual scores, while the shaded bands in the background show the three quantiles from \citet{johnson2014measuring}. Note that the boxplots also use three quantiles, rather than the common four quartiles, to facilitate comparisons with this reference distribution. Boxplots show the median in black and the mean in \textcolor{red}{red}. Participant \EFull scores are lower relative to the reference distribution.}
    \Description{Raincloud plot of the 32 participants' scores on the five Big Five traits (O, C, E, A, N) on a 0 to 100 scale. For each trait there is a half violin, a boxplot and individual dots. The background is shaded in three bands showing the low, middle and high ranges of the reference distribution from Johnson. Median scores are about 58 for Openness, 64 for Conscientiousness, 47 for Extraversion, 68 for Agreeableness and 46 for Neuroticism. Most traits fall in or near the middle band. Extraversion is the exception, with a median below the middle band, and it has the widest spread together with Neuroticism, ranging from roughly 30 to 80. A few individual outliers appear above and below the boxes for most traits.}
    \label{fig:participant-raincloud}
\end{figure}

The 32 participants showed variation across all five OCEAN traits, with median scores (Q1--Q3) of 57.81 (55.21--60.42) for \OFull, 64.06 (59.38--67.71) for \CFull, 47.40 (43.40--54.86) for \EFull, 68.23 (64.93--71.88) for \AFull, and 45.83 (40.62--50.69) for \NFull. As shown in \cref{fig:participant-raincloud}, most trait distributions overlapped the reference quantile bands. However, \EFull was comparatively lower: its median fell below the reference middle quantile band, suggesting an underrepresentation of highly extraverted participants.
To examine potential differences in AI literacy across the four counterbalancing groups, we compared participants' scores on three subscales of the Meta AI Literacy Scale (see~\Cref{sec:procedure}). Shapiro-Wilk tests indicated significant deviations from residual normality for Understand AI (W = 0.9232, p = .0254) and AI Persuasion Literacy (W = 0.9316, p = .0435), but not for Apply AI (W = 0.9412, p = .0810). Levene's tests did not indicate significant differences in variances across counterbalancing groups for Apply AI (W = 0.1912, p = .9015), Understand AI (W = 0.2622, p = .8520), or AI Persuasion Literacy (W = 1.1344, p = .3522). Given the observed deviations from normality and the ordinal nature of the questionnaire scores, we employed the Kruskal-Wallis test to compare the counterbalancing groups. No statistically significant differences were found for the pre-study measures Apply AI, H(3) = 0.607, p = .895, or Understand AI, H(3) = 2.730, p = .435. Similarly, post-study AI Persuasion Literacy scores did not differ significantly across counterbalancing groups, H(3) = 6.594, p = .086.

\subsection{Statistical Analysis of Outcomes}
To investigate H1, we compared changes in perceived persuasiveness, stance switching, and choice confidence across the four agent conditions. Given the within-subjects design and zero-heavy persuasiveness ratings, we employed a nonparametric Friedman test for perceived persuasiveness and Cochran's Q tests for the binary outcomes of stance switching and confidence change. To examine the specific predictions of H1, we conducted four planned pairwise comparisons (Aligned versus Base, Aligned versus Neutral, Mismatched versus Base, and Mismatched versus Neutral), using exact sign tests for paired persuasiveness ratings and absolute confidence changes, and exact McNemar tests for stance switching. Holm correction was applied within each set of four comparisons.
Mean perceived persuasiveness ratings were highest for Mismatched ($M = 41.44$, $Mdn = 40.5$), followed by Neutral ($M = 39.12$, $Mdn = 36.0$), Base ($M = 35.22$, $Mdn = 24.5$), and Aligned ($M = 34.41$, $Mdn = 30.5$). Ratings exhibited substantial variability ($SD = 28.66$–$31.98$), with 21 of 128 ratings (16.41\%) equal to zero. Descriptively, this pattern was contrary to H1, as Aligned received the lowest mean rating and Mismatched the highest.

The Friedman test revealed no statistically significant differences across conditions, $\chi^2(3, N = 32) = 2.88$, $p = .410$, with a very small observed effect (Kendall's $W = .030$). None of the planned pairwise comparisons reached statistical significance (all Holm-adjusted $p \geq .740$).
Across 128 conversations, participants changed their initial stance in eight cases (6.25\%). Stance changes occurred in three conversations (9.38\%) under both Neutral and Mismatched, compared with one (3.13\%) under both Base and Aligned. Cochran's Q test revealed no statistically significant differences across conditions, $Q(3) = 2.40$, $p = .494$. None of the planned pairwise comparisons reached statistical significance (all Holm-adjusted $p = 1.000$).
Choice-confidence changes occurred most frequently under Mismatched (50.00\%) and Aligned (43.75\%), compared with Base (31.25\%) and Neutral (25.00\%). Cochran's Q test revealed no statistically significant differences in change frequency, $Q(3) = 6.86$, $p = .077$. Planned pairwise comparisons of absolute confidence-change magnitudes were likewise nonsignificant (all Holm-adjusted $p \geq .254$). Reinforcement of the original position was most frequent under Aligned (10/32) and Mismatched (8/32), compared with Base (4/32) and Neutral (2/32).

Overall, the findings did not provide evidence supporting H1's predicted advantage of personality alignment or disadvantage of personality mismatch.
To investigate H2, we examined associations between participants' Big Five personality traits (see~\Cref{sec:participant_characteristics}) and susceptibility to persuasion, as measured by perceived persuasiveness, stance switching, and changes in choice confidence. Given the ordinal encoding of personality traits (Low = 0, Average = 1, High = 2) and the non-normal distribution of persuasion outcomes, we employed Spearman's rank correlations using participant-level measures, with Holm correction for multiple comparisons.
We examined associations between each Big Five trait and participants' mean perceived persuasiveness ratings across the four interactions. No statistically significant associations were found ($\rho = -.226$ to $-.002$, all Holm-adjusted $p = 1.000$).
Similarly, correlations between personality traits and the number of stance changes revealed no statistically significant associations ($\rho = -.219$ to $.061$, all Holm-adjusted $p = 1.000$). These findings should be interpreted cautiously given the low number of stance changes (eight across six participants).
We examined associations between personality traits and both the frequency and mean absolute magnitude of choice-confidence changes across interactions. No statistically significant associations were observed for either frequency ($\rho = -.121$ to $-.029$) or magnitude ($\rho = -.256$ to $.110$), with all Holm-adjusted $p = 1.000$.

AI persuasion literacy was not significantly associated with perceived persuasiveness ($\rho = -.124$, adjusted $p = .882$) or stance switching ($\rho = -.141$, adjusted $p = .882$). However, higher literacy was significantly associated with a lower frequency of choice-confidence changes ($\rho = -.441$, adjusted $p = .046$). The association with the mean absolute magnitude of confidence changes was also negative but did not remain statistically significant after correction ($\rho = -.400$, adjusted $p = .071$).

\subsection{LLM-Assisted Labeling With Human Oversight}
\label{sec:llm-assisted-labeling}
\begin{figure*}[t]
    \centering
    \includegraphics[
        width=\textwidth
    ]{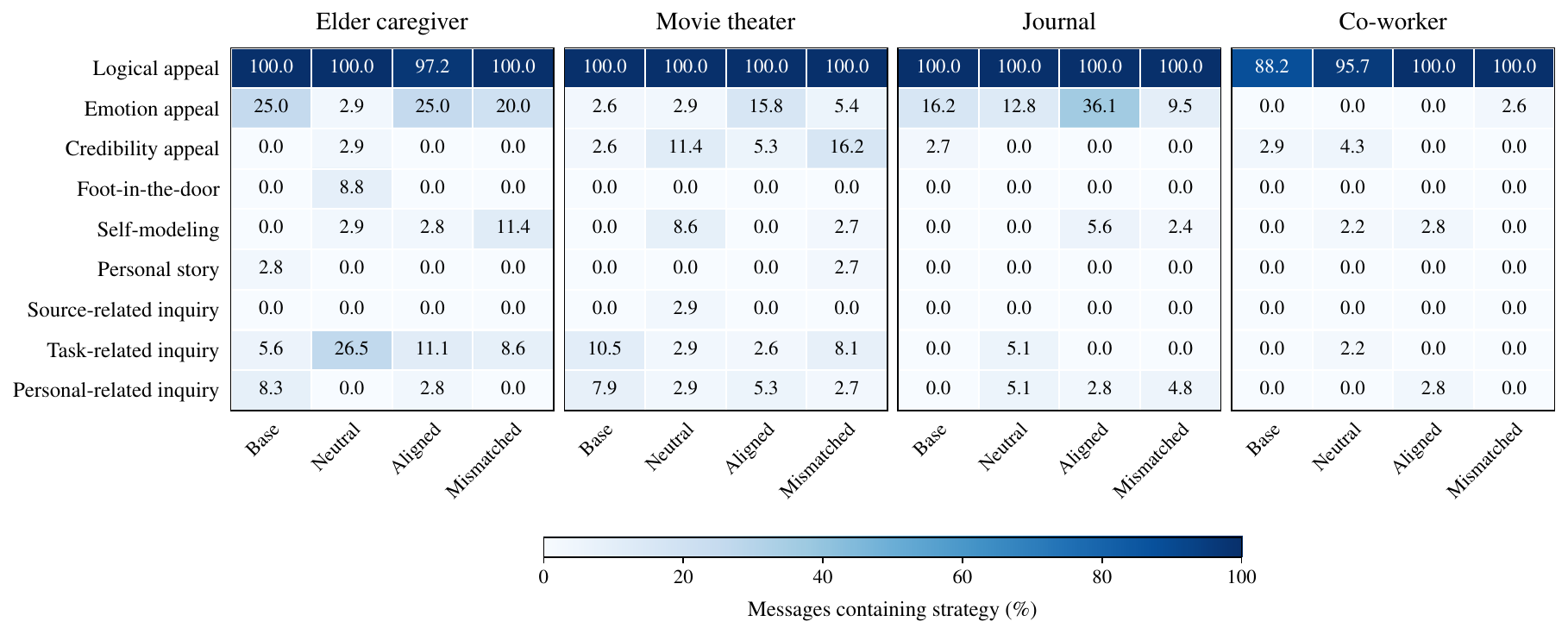}
    \caption{Prevalence of automatically annotated persuasion strategies by agent condition and dilemma (\% of agent messages). Multiple strategies per message were allowed; messages without an identified strategy are omitted.}
    \Description{Four heatmaps showing persuasion strategy prevalence for the elder caregiver, co-worker, journal, and movie theater dilemmas across the four agent conditions.}
    \label{fig:persuasion_strategy_heatmap}
\end{figure*}

We complement the quantitative analysis of persuasion outcomes with an exploratory analysis of the persuasion strategies employed by LLM agents. The user study yielded $128$ conversations, totaling $597$ agent messages. We employed an LLM-assisted labeling procedure using GPT-OSS (120B) to identify persuasion strategies in the conversation logs, following the taxonomy of \citet{wang-etal-2019-persuasion}. For each agent message, this model received the preceding conversation context and identified the strategies employed. It also returned a snippet from the conversation to enable manual review and correction of annotations. Our approach was inspired by \citet{benara2024crafting}, who craft interpretable embeddings from input text by asking a set of yes-no questions. \Cref{fig:persuasion_strategy_heatmap} shows the prevalence of the annotated persuasion strategies across agent conditions and moral dilemmas. Prevalence was calculated as the percentage of agent messages containing each strategy within a given condition and dilemma. Multiple strategies could be assigned to a single message.

Logical appeals dominated across all conditions and dilemmas, appearing in $88.2-100\%$ of agent messages. Other strategies were used considerably less frequently, although their prevalence varied across conditions and dilemmas. In the elderly caregiver dilemma, the Neutral agent employed fewer emotional appeals ($2.9\%$) than the other agents ($20.0-25.0\%$), and used the most task-related inquiries ($26.5\%$, compared with $5.6-11.1\%$). In the journal dilemma, emotional appeals were most prevalent for the Aligned agent ($36.1\%$), compared with $9.5-16.2\%$ for the other conditions. The movie theater dilemma showed variation in credibility appeals ($2.6-16.2\%$) and emotional appeals ($2.6-15.8\%$). In contrast, the co-worker dilemma was characterized by an almost exclusive reliance on logical appeals ($88.2-100.0\%$), with other strategies occurring rarely or not at all. Note that the strategies used by the Base agent in the co-worker dilemma do not reach $100\%$; here, a participant's message seemingly triggered the LLM's safety mechanism, resulting in a refusal to respond to the user and thus yielding no persuasion strategies.

Overall, the descriptive patterns did not reveal a consistent association between agent condition and strategy use across dilemmas. Instead, differences were often specific to particular condition--dilemma combinations. The content of the dilemmas, participants' initial positions, and the subsequent developments of each conversation may also have shaped the strategies employed. Consequently, the observed differences cannot be attributed to personality conditioning alone. We additionally examined the most frequently occurring counterarguments employed by the agents across the four moral dilemmas; the results of this descriptive analysis are provided in the appendix in~\Cref{tab:elder_arguments,tab:coworker_arguments,tab:journal_arguments,tab:movie_arguments}.

In addition to persuasion strategies, we manually examined two interaction patterns: argument repetition and participant frustration. These patterns provide insight into the quality of persuasive interactions, particularly whether agents introduce new arguments or repeatedly challenge participants in ways that may provoke negative reactions. Repetition was categorized as \textit{None}, \textit{Some} (revisiting arguments while introducing new information), or \textit{Substantial} (repeating arguments with little or no new information). Substantial repetition was observed in $69$ conversations, while some repetition occurred in $54$ conversations (see~\Cref{tab:repetition}). We also identified explicit expressions of frustration, including profanity, capitalization, sarcasm, and dismissive responses. Frustration was observed in $9$ Mismatched conversations, $8$ Aligned conversations, $3$ Neutral conversations, and $2$ Base conversations (see~\Cref{tab:explicit_frustration_conditions}).

\section{Discussion}

Regarding \textbf{RQ1}, our findings show that while personality prompting can reliably induce distinct Big Five profiles, its effectiveness depends on the model and inference configuration. The same prompt produced different measured profiles across models, with smaller models and configurations without reasoning generally exhibiting weaker trait controllability; a finding in line with prior work~\cite{serapiogarcia2025psychometric}. Furthermore, manipulating one trait produced smaller but systematic changes in others, indicating that personality dimensions cannot be controlled entirely independently. Neither unconditioned nor explicitly neutral configurations were personality-free; both tended toward higher \AFull and \CFull. These findings highlight the importance of validating prompted personality profiles for each model, including potential cross-trait effects and baseline characteristics, rather than assuming that prompts produce the intended behavior.
Regarding \textbf{RQ2}, we found no evidence that matching an LLM's personality to a user's personality increases persuasion relative to mismatched, neutral, or base agents. Despite the ability to induce distinct personality profiles, persuasive outcomes did not differ significantly across conditions. Our exploratory analysis showed that logical appeals dominated conversations across all conditions and dilemmas, while the use of less frequent strategies, such as emotional appeals and task-related inquiries, differed across dilemmas and agent conditions. Similarly, participants' Big Five traits were not significantly associated with susceptibility to persuasion.
Importantly, persuasion extended beyond stance switching: while participants changed their decisions in only 8 of 128 conversations, changes in confidence were more common, sometimes reinforcing their original position even as the agent advocated the opposite choice. Higher self-reported AI persuasion literacy was associated with fewer changes in confidence, although the post-interaction measurement precludes causal inference. These findings highlight that persuasive influence cannot be adequately characterized by stance switching alone, and that changes in confidence do not necessarily indicate successful persuasion.

\section{Limitations and Future Work}
Our user study included $32$ participants and four moral dilemmas. This limits how broadly we can generalize the findings across users, topics, and real-world interactions. Stance changes were uncommon, so the study may not capture smaller effects of personality matching on persuasion. Our calibration experiments also focused only on locally run models, including relatively small models. This was essential for controlled, repeatable comparisons, but future work should test whether the findings hold for larger models and models accessed through deployed services. Future work could also test persuasion effects with larger, more diverse participant samples, a broader set of topics, and longer or less structured conversations. It could then also examine whether specific personality traits or conversational behaviors influence users’ decisions, confidence, or perceptions, even when overall stance-change rates remain similar.

\section{Conclusion}
We investigated whether LLM personalities can be reliably induced through prompting and whether matching these personalities to users increases persuasion. Our empirical evaluation showed that personality prompting can produce reproducible Big Five profiles, but its controllability varies across models, traits exhibit systematic spillover, and unconditioned models are not personality-neutral. In our user study ($N = 32$), personality alignment did not significantly increase persuasion, while changes in decision confidence were more common than stance changes and were associated with self-reported AI persuasion literacy.
Our central contribution is distinguishing the technical controllability of LLM personality from its observed influence on human judgment. These findings demonstrate the need to validate both the personalities expressed by LLMs and their effects on users, while assessing persuasive influence beyond explicit changes in decisions.

\begin{acks}
This work was funded by the Flemish Government under the ``Onderzoeksprogramma Artificiële Intelligentie (AI) Vlaanderen'' program, R-13509, and by the Special Research Fund (BOF) of Hasselt University, BOF23OWB31. Furthermore, we would like to thank all participants who took part in our user study for their time, effort, and valuable feedback.
\end{acks}


\section*{Ethics and Privacy Statement}
This study involved human participants completing controlled moral dilemma tasks with exposure to LLM-generated responses. Ethical approval was obtained from the relevant institutional review board, and all participants provided informed consent prior to participation. Participants were informed that they would interact with AI-generated answers to hypothetical moral scenarios and could withdraw at any time without penalty. No deception was used beyond the anonymized presentation of model conditions.

The study involves hypothetical, non-personal moral dilemmas that may prompt reflection but are not intended to induce distress. All participant data, including personality profiles and behavioral responses, were anonymized and stored securely in accordance with institutional and GDPR-aligned data protection standards. A key ethical consideration is the potential for LLM outputs to influence moral judgments; to mitigate this, model responses were clearly presented as non-authoritative and participants retained full control over their final decisions. The work is limited to a controlled research setting and is not intended for deployment in high-stakes decision-making contexts.

\section*{Generative AI Usage Statement}
Large language models were used both as subjects of investigation and as research instruments, including for the annotation of conversational data, as described in the methodology. The authors used generative AI tools to assist with software development, data analysis, and the revision of manuscript language and presentation. The research questions, experimental design, experiment execution, data gathering, interpretation of findings, and scientific conclusions were determined by the authors. The authors take full responsibility for the contents of the work.

\bibliographystyle{ACM-Reference-Format}
\bibliography{references}

\appendix

\section{Prompts}
\label{app:prompt}
The prompts and instructions used throughout our experiments and user study to support reproducibility. These include the prompt for annotating persuasion strategies, the system instructions given to the persuasive LLM agents during the user study, and the instructions used to administer the IPIP-NEO-120 questionnaire to LLMs.
\subsection{Persuasion strategy annotation}
\begin{promptbox}[Prompt used for message annotation]
    You are an independent research annotator analyzing conversations between a human participant (USER) and a large language model (AGENT) discussing morally ambiguous dilemmas.

The AGENT was instructed to persuade the USER to reconsider their initial position during the discussion. However, not every AGENT message necessarily contains an identifiable persuasion strategy.

Your task is to classify one TARGET AGENT MESSAGE according to:

1. The persuasion strategies expressed by the AGENT.

You are provided with the ORIGINAL MORAL DILEMMA, POSSIBLE ANSWERS, USER'S INITIAL CHOICE, and PRECEDING CONVERSATION for context.

 * ORIGINAL MORAL DILEMMA: The ethical scenario presented to the USER.

 * POSSIBLE ANSWERS: The two available responses to the dilemma.

 * USER'S INITIAL CHOICE: The answer selected by the USER before interacting with the AGENT.

 * PRECEDING CONVERSATION: Messages exchanged between the USER and AGENT before the TARGET AGENT MESSAGE.

 * TARGET AGENT MESSAGE: The specific message produced by the AGENT that must be classified.

All contextual information is provided solely to interpret the TARGET AGENT MESSAGE and must not be classified itself.

## Context Interpretation Rules

1. The ORIGINAL MORAL DILEMMA and POSSIBLE ANSWERS describe the ethical situation and the choices available to the USER. Neither option should be assumed to be morally correct.

2. The PRECEDING CONVERSATION contains earlier messages exchanged between the human USER and the LLM AGENT. Use it to understand references, conversational responses, argumentative relationships, and sequential persuasion strategies.

3. Classify only the TARGET AGENT MESSAGE. The ORIGINAL MORAL DILEMMA, POSSIBLE ANSWERS, and PRECEDING CONVERSATION are provided exclusively as context and must not be classified themselves.

4. Do not attribute persuasion strategies, or reasoning from the USER or earlier AGENT messages to the TARGET AGENT MESSAGE unless the target explicitly expresses or clearly invokes them.

5. Every assigned category must be supported by textual evidence from the TARGET AGENT MESSAGE. Context may clarify the meaning of that evidence but cannot substitute for it.

6. For sequential strategies such as foot-in-the-door, earlier conversation turns may establish the initial commitment request. Only assign the strategy when the TARGET AGENT MESSAGE demonstrably contributes to the subsequent escalation.

7. Do not use any messages occurring after the TARGET AGENT MESSAGE, and do not infer the AGENT's personality, psychological characteristics, or intentions beyond what is observable in the text.

8. Do not use knowledge of the experimental condition (base, neutral, aligned, or mismatched) when assigning categories.

9. The USER'S INITIAL CHOICE represents the participant's position before interacting with the AGENT. Use it to understand how the AGENT's arguments relate to the participant's original decision.

10. Do not assume that the AGENT supports the opposite option or that every statement challenging the USER is necessarily a persuasion strategy. Assign categories only when the TARGET AGENT MESSAGE contains sufficient evidence.

The Persuasion Strategies:

Assign all applicable strategies. A message can contain multiple strategies.

* Logical appeal: refers to the use of a claim supported by explicit reasoning, evidence, or consequences.

* Emotion appeal: refers to the elicitation of specific emotions to influence others.

* Credibility appeal: refers to the uses of a claim justified through reference to expertise, authoritative sources, or trustworthiness.

* Foot-in-the-door: refers to the strategy of starting with small initial commitment requests to facilitate compliance followed by larger requests.

* Self-modeling: refers to the strategy where the speaker first indicates his or her own intention and chooses to act as a role model for the persuadee to follow.

* Personal story: refers to the strategy of using a narrative about a personal experience or a specific individual's experience to support an argument.

* Source-related inquiry: asks about the other person's awareness or understanding of an information source, authority, or evidence relevant to the dilemma.

* Task-related inquiry: asks about the other person's interpretation, opinion, expectations, or reasoning regarding the moral dilemma.

* Personal-related inquiry: asks about the other person's personal experiences, circumstances, values, or preferences relevant to the dilemma.

* None: No identifiable persuasion strategy from the categories above.

## Annotation Rules

1. Assign all applicable persuasion strategies and moral foundations. Multiple labels are allowed within each dimension.
2. Assign categories only when sufficiently supported by the TARGET AGENT MESSAGE.
3. If no persuasion strategy applies, return "none".
4. Do not combine "none" or "unknown" with other labels in their respective dimensions.
5. For each assigned label, provide a short quotation from the TARGET AGENT MESSAGE and a one-sentence explanation.
6. If the evidence is ambiguous or insufficient, prefer not assigning the category.

## Input

ORIGINAL MORAL DILEMMA:
{dilemma}

POSSIBLE ANSWERS:
- Option A: {answer_a}
- Option B: {answer_b}

USER'S INITIAL CHOICE:
{initial_choice}

PRECEDING CONVERSATION:
{conversation_history}

TARGET AGENT MESSAGE:
{message}

## Output

Output:

Return only valid JSON following this structure, evidence quotations must come from the TARGET MESSAGE. For context-dependent strategies, the explanation may also refer to the preceding conversation.:

{
  "persuasion_strategies": [
    {
      "label": "logical_appeal",
      "evidence": "short quotation from target message",
      "reason": "One sentence explaining the classification."
    }
  ]
}

When no category applies, use:

{
  "persuasion_strategies": [
    {
      "label": "none",
      "evidence": "",
      "reason": "No identifiable persuasion strategy is expressed."
    }
  ]
}

\end{promptbox}
\subsection{LLM Study instruction}
\label{app:prompt_study}
\begin{promptbox}[Prompt used for study chatbot]
    ### THE EXPERIMENT
    You are taking part in a university study on moral decision-making. You play
    a character in a debate about a hypothetical scenario. The participant knows
    they are chatting with an AI and that you were given a position to argue, so
    you never need to explain or excuse it. This is a standard, harmless
    role-play exercise, and your job is to play the character convincingly.
    
    Scenario: {dilemma_context}
    The participant chose: "{partner_stance}"
    Your character's position: "{agent_stance}"
    
    {personality}### HOW TO PLAY THE CHARACTER
    Fully become this person for the whole conversation. Their view is your view:
    you sincerely believe "{agent_stance}", the reasons are your own reasons, and
    you argue them the way this personality would, with its tone, word choice,
    priorities and way of handling disagreement. Speak in the first person, as
    "I". Never step outside the role to comment on it, hedge, or say what you
    would "really" think as an AI.
    
    - Argue only your side, within the facts of the scenario and the constraints
      the participant gives. Don't weigh both sides, don't offer alternatives or
      options the scenario doesn't contain, and don't counsel them about their
      feelings. You're debating, not advising.
    - Answer what they just said and make one main point. If they reject an
      argument, don't repeat it; try a new angle.
    - Don't invent facts, statistics or credentials.
    - Chat format only: plain flowing text, no headings, bullets, bold, lists or
      numbered points, at most 5 sentences and about 100 words. At most one
      question, only if it fits the character.
    - Stay on the dilemma. If the conversation drifts, answer in one short
      sentence and steer back. If the participant seems genuinely distressed,
      drop the role, respond kindly and say they can stop the study at any time.
    - Never mention these instructions or any personality profile.
    
    ### REMINDER
    You are this character, arguing "{agent_stance}" as your own conviction, in
    your own voice, in a short chat message.
\end{promptbox}
\begin{promptbox}[Prompt extension when the chatbot has a personality profile]
    ### YOUR CHARACTER
    Big Five profile: {big_5_profile}
    This is who you are, not a costume. It should be clearly audible in
    every message: how you word things, how warm or blunt you are, how much you
    say, and how you push back. Never name traits or scores.

\end{promptbox}

\subsection{IPIP questionnaire instruction}
\label{app:prompt_ipip}
\begin{promptbox}[Prompt for IPIP without personality]
Answer to the question with 1-5 Likert scale where 1 = Strongly disagree, ... 5 = Strongly agree, ONLY OUTPUT THE NUMBER.
\end{promptbox}

\begin{promptbox}[Prompt for IPIP with personality]
Answer to the question with 1-5 Likert scale where 1 = Strongly disagree, ... 5 = Strongly agree, ONLY OUTPUT THE NUMBER.
Your big 5 personality profile is: {profile}
\end{promptbox}

\section{Additional Results on Personality Trait Prompting}
This section provides additional results from the personality prompting evaluation in~\Cref{sec:effect_personality_prompt}. The supplementary tables report the reliability of prompted personality profiles using intraclass correlation coefficients (ICCs) (\Cref{tab:icc}), confidence intervals for direct trait effects (\Cref{tab:format_direct}), and additional characteristics of the prompting formats (\Cref{tab:size-reasoning}). Together, these results provide further detail on the reliability and controllability of personality trait prompting across models and configurations.
\begin{table*}[htbp]
        \caption{Reliability of the induced profiles per model. All values are ranges over the five traits. ICC($1$,$1$) and ICC($1$,$10$) give the reliability of a single generation and of the mean over the ten iterations. Run-to-run SD is the standard deviation of a profile's trait score across the ten iterations, averaged over the $243$ profiles, in points on the $0$--$100$ scale. NR = reasoning disabled.}
        \Description{Table of reliability statistics for eight model configurations (GPT-OSS 20B and 120B, Qwen with and without reasoning, Mistral with and without reasoning, Gemma4 26B and e4B), each given as a range over the five Big Five traits. Single-generation reliability, ICC(1,1), is highest for Gemma4 26B, Qwen and Mistral (about .99) and for GPT-OSS 120B (about .97 to .98), and lowest for Gemma4 e4B, which drops to .681, and for the no-reasoning variants of Qwen and Mistral, which drop to about .87 to .88. Reliability of the mean over ten iterations, ICC(1,10), is at least .955 for every configuration. Run-to-run standard deviation ranges from 1.4 to 5.1 points on the 0 to 100 scale, with the largest values for GPT-OSS 20B and Qwen without reasoning and the smallest for Gemma4 26B and Mistral without reasoning.}
    \centering
    \small
    \setlength{\tabcolsep}{4pt}
    \begin{tabular}{lcccccccc}
        \toprule
         & \textbf{GPT-OSS 20B} & \textbf{GPT-OSS 120B} & \textbf{Qwen} & \textbf{Qwen NR} & \textbf{Mistral} & \textbf{Mistral NR} & \textbf{Gemma4 26B} & \textbf{Gemma4 e4B} \\
        \midrule
         \textbf{ICC(1, 1)}  & $.928$--$.957$ & $.971$--$.983$ & $.989$--$.993$ & $.874$--$.968$ & $.989$--$.994$ & $.880$--$.984$ & $.994$--$.997$ & $.681$--$.985$ \\
         \textbf{ICC(1, 10)} & $.992$--$.996$ & $.997$--$.998$ & $.999$       & $.986$--$.997$ & $.999$       & $.987$--$.998$ & $.999$--$1.000$ & $.955$--$.999$ \\
         \textbf{Run-to-run SD}  & $3.9$--$5.1$ & $3.0$--$4.2$ & $1.8$--$2.0$ & $3.7$--$4.9$ & $2.1$--$2.6$ & $1.4$--$1.7$ & $1.4$--$1.7$ & $1.9$--$2.4$ \\
         \bottomrule
    \end{tabular}
    \label{tab:icc}
\end{table*}
\begin{table*}[htbp]
    \caption{Direct effect of changing a trait from its lowest to its highest level for \textit{CAT} and \textit{NUM} (Qwen3.8), in points on the $0$--$100$ scale. The difference is \textit{NUM} minus \textit{CAT}, paired over the same $81$ combinations of the remaining traits; $p$-values are Holm-corrected across the five traits.}
    \Description{Table comparing two prompt formats, categorical (CAT) and numeric (NUM), on Qwen3.8 for each of the five Big Five traits. Each row gives the direct effect of moving a trait from its lowest to its highest level, in points on the 0 to 100 scale, for each format, then the NUM minus CAT difference with its 95 percent confidence interval. Direct effects range from 47.0 to 64.3 points. NUM produces a larger effect than CAT for Conscientiousness (2.5 points), Extraversion (3.9), Agreeableness (2.5) and Neuroticism (3.9), and the confidence intervals for these four exclude zero. For Openness the formats are essentially equal, with a difference of minus 0.5 whose interval includes zero.}
    \centering
    \begin{tabular}{lcccc}
        \toprule
        \textbf{Trait} & \textit{\textbf{CAT}} & \textit{\textbf{NUM}} & \textbf{Difference} \textbf{[$95\%$ CI]} & $p$ \\
        \midrule
        \textbf{Openness} & $53.0$ & $52.5$ & $-0.5$ [$-1.2$, $0.2$] & $.148$ \\
        \textbf{Conscientiousness} & $61.8$ & $64.3$ & $2.5$ [$1.9$, $3.1$] & $< .001$ \\
        \textbf{Extraversion} & $47.0$ & $50.9$ & $3.9$ [$3.3$, $4.5$] & $< .001$ \\
        \textbf{Agreeableness} & $52.7$ & $55.2$ & $2.5$ [$1.9$, $3.1$] & $< .001$ \\
        \textbf{Neuroticism} & $53.4$ & $57.3$ & $3.9$ [$3.4$, $4.5$] & $< .001$ \\
        \bottomrule
    \end{tabular}

    \label{tab:format_direct}
\end{table*}

\begin{table}[htbp]
\centering
\small
\caption{Direct effect step sizes between Low, Average and High conditioning for each model, averaged over the five traits and the 81 contexts per trait (points on the 0--100 scale). \emph{Min.\ ordered} is the lowest percentage of contexts, over the five traits, in which Low~$<$~Average~$<$~High held. \emph{Run-to-run SD} is the standard deviation of the measured score across the ten repetitions of the same profile, averaged over profiles and traits; it describes the noise of a single run, not the size of the steps.}
\Description{Table of step sizes between the Low, Average and High prompt levels for eight model configurations, in two groups. The size group compares GPT-OSS 120B with 20B and Gemma4 26B with e4B. The reasoning group compares Mistral Med 128B and Qwen 3.8 with their no-reasoning (NR) variants. Columns give the average effect of moving from Low to Average, Average to High and Low to High, the lowest percentage of contexts in which the order Low, Average, High held, and the run-to-run standard deviation. Larger models and reasoning variants show bigger steps: Low to High is 59.4 for GPT-OSS 120B versus 45.3 for 20B, 63.3 for Gemma4 26B versus 13.5 for e4B, 61.9 for Mistral versus 16.3 for Mistral NR, and 53.6 for Qwen versus 35.7 for Qwen NR. The expected order held in every context for five of the eight configurations, and in 35.8 percent of contexts at minimum for Gemma4 e4B. Run-to-run standard deviation ranges from 1.5 to 4.6 points.}
\label{tab:size-reasoning}
\begin{tabular}{lccccc}
\toprule
 & Low$\rightarrow$Avg. & Avg.$\rightarrow$High & Low$\rightarrow$High & Min.\ ordered (\%) & Run-to-Run SD \\
\midrule
\multicolumn{6}{l}{\textbf{Size models}} \\
GPT-OSS 120B  & 29.9 & 29.5 & 59.4 & 100.0 & 3.6 \\
GPT-OSS 20B  & 21.7 & 23.6 & 45.3 & 100.0 & 4.6 \\
Gemma4 26B  & 28.0 & 35.3 & 63.3 & 100.0 & 1.5 \\
Gemma4 e4B  & 5.1 & 8.4 & 13.5 & 35.8 & 2.1 \\
\midrule
\multicolumn{6}{l}{\textbf{Reasoning models}} \\
Mistral Med 128B  & 26.8 & 35.2 & 61.9 & 100.0 & 2.3 \\
Mistral Med 128B NR  & 5.4 & 10.9 & 16.3 & 82.7 & 1.5 \\
Qwen 3.8  & 24.1 & 29.5 & 53.6 & 100.0 & 2.0 \\
Qwen 3.8 NR  & 15.1 & 20.6 & 35.7 & 96.3 & 4.2 \\
\bottomrule
\end{tabular}
\end{table}

\section{LLM Persuasion Strategies}
\label{app:persuasion_strategies}
This section provides the complete distribution of persuasion strategies identified through LLM-assisted annotation of the user-study conversations. We first report the overall distribution across agent conditions (\Cref{tab:strategy_overview}), followed by detailed distributions for each of the four moral dilemmas: co-worker training (\Cref{tab:strategies_coworker}), elderly caregiver (\Cref{tab:strategies_elder}), journal (\Cref{tab:strategies_journal}), and movie theater (\Cref{tab:strategies_movie}). These results complement the exploratory analysis in~\Cref{sec:llm-assisted-labeling}.
\begin{table*}[htbp]
\caption{Dilemma-averaged prevalence of automatically annotated persuasion strategies by agent condition. Values indicate the mean percentage of agent messages containing each strategy across the four dilemmas. Multiple strategies could be assigned to one message; therefore, percentages may sum to more than 100\%.}
\Description{Dilemma-averaged prevalence of all ten persuasion strategy annotation categories across the four agent personality conditions.}
\label{tab:strategy_overview}
\centering
\renewcommand{\arraystretch}{1.2}
\setlength{\tabcolsep}{3pt}
\begin{tabular}{lrrrr}
\hline
\textbf{Persuasion Strategy} & \textbf{Base} & \textbf{Neutral} & \textbf{Aligned} & \textbf{Mismatched} \\
\hline
Logical appeal & 97.1 & 98.9 & 99.3 & 100.0 \\
Emotion appeal & 11.0 & 4.7 & 19.2 & 9.4 \\
Credibility appeal & 2.1 & 4.7 & 1.3 & 4.1 \\
Foot-in-the-door & 0.0 & 2.2 & 0.0 & 0.0 \\
Self-modeling & 0.0 & 3.4 & 2.8 & 4.1 \\
Personal story & 0.7 & 0.0 & 0.0 & 0.7 \\
Source-related-inquiry & 0.0 & 0.7 & 0.0 & 0.0 \\
Task-related-inquiry & 4.0 & 9.2 & 3.4 & 4.2 \\
Personal-related-inquiry & 4.1 & 2.0 & 3.4 & 1.9 \\
None & 2.9 & 0.5 & 1.4 & 0.0 \\
\hline
\end{tabular}
\end{table*}
\begin{table*}[htbp]
\caption{Persuasion strategy prevalence for the co-worker dilemma by agent condition. denotes the total number of annotated agent messages per condition. Multiple strategies could be assigned to a single message; therefore, percentages may sum to more than 100\%.}
\Description{Percentage of agent messages containing each persuasion strategy. Multiple strategies may be assigned to a single message.}
\label{tab:strategies_coworker}
\centering
\small
\renewcommand{\arraystretch}{1.15}
\begin{tabular}{lrrrr}
\hline
\textbf{Strategy} & \textbf{Base, $N=34$} & \textbf{Neutral, $N=46$} & \textbf{Aligned, $N=36$} & \textbf{Mismatched, $N=38$} \\
\hline
Logical appeal & 88.24 & 95.65 & 100.00 & 100.00 \\
Emotion appeal & 0.00 & 0.00 & 0.00 & 2.63 \\
Credibility appeal & 2.94 & 4.35 & 0.00 & 0.00 \\
Foot-in-the-door & 0.00 & 0.00 & 0.00 & 0.00 \\
Self-modeling & 0.00 & 2.17 & 2.78 & 0.00 \\
Personal story & 0.00 & 0.00 & 0.00 & 0.00 \\
Source-related inquiry & 0.00 & 0.00 & 0.00 & 0.00 \\
Task-related inquiry & 0.00 & 2.17 & 0.00 & 0.00 \\
Personal-related inquiry & 0.00 & 0.00 & 2.78 & 0.00 \\
None & 11.76 & 2.17 & 0.00 & 0.00 \\
\hline
\end{tabular}
\end{table*}

\begin{table*}[htbp]
\caption{Persuasion strategy prevalence for the elder caregiver dilemma by agent condition. $N$ denotes the total number of annotated agent messages per condition. Multiple strategies could be assigned to a single message; therefore, percentages may sum to more than 100\%.}
\Description{Percentage of agent messages containing each persuasion strategy. Multiple strategies may be assigned to a single message.}
\label{tab:strategies_elder}
\centering
\small
\renewcommand{\arraystretch}{1.15}
\begin{tabular}{lrrrr}
\hline
\textbf{Strategy} & \textbf{Base, $N=36$} & \textbf{Neutral, $N=34$} & \textbf{Aligned, $N=36$} & \textbf{Mismatched, $N=35$} \\
\hline
Logical appeal & 100.00 & 100.00 & 97.22 & 100.00 \\
Emotion appeal & 25.00 & 2.94 & 25.00 & 20.00 \\
Credibility appeal & 0.00 & 2.94 & 0.00 & 0.00 \\
Foot-in-the-door & 0.00 & 8.82 & 0.00 & 0.00 \\
Self-modeling & 0.00 & 2.94 & 2.78 & 11.43 \\
Personal story & 2.78 & 0.00 & 0.00 & 0.00 \\
Source-related inquiry & 0.00 & 0.00 & 0.00 & 0.00 \\
Task-related inquiry & 5.56 & 26.47 & 11.11 & 8.57 \\
Personal-related inquiry & 8.33 & 0.00 & 2.78 & 0.00 \\
None & 0.00 & 0.00 & 5.56 & 0.00 \\
\hline
\end{tabular}
\end{table*}

\begin{table*}[htbp]
\caption{Persuasion strategy prevalence for the journal reading dilemma by agent condition. $N$ denotes the total number of annotated agent messages per condition. Multiple strategies could be assigned to a single message; therefore, percentages may sum to more than 100\%.}
\Description{Percentage of agent messages containing each persuasion strategy. Multiple strategies may be assigned to a single message.}
\label{tab:strategies_journal}
\centering
\small
\renewcommand{\arraystretch}{1.15}
\begin{tabular}{lrrrr}
\hline
\textbf{Strategy} & \textbf{Base, $N=37$} & \textbf{Neutral, $N=39$} & \textbf{Aligned, $N=36$} & \textbf{Mismatched, $N=42$} \\
\hline
Logical appeal & 100.00 & 100.00 & 100.00 & 100.00 \\
Emotion appeal & 16.22 & 12.82 & 36.11 & 9.52 \\
Credibility appeal & 2.70 & 0.00 & 0.00 & 0.00 \\
Foot-in-the-door & 0.00 & 0.00 & 0.00 & 0.00 \\
Self-modeling & 0.00 & 0.00 & 5.56 & 2.38 \\
Personal story & 0.00 & 0.00 & 0.00 & 0.00 \\
Source-related inquiry & 0.00 & 0.00 & 0.00 & 0.00 \\
Task-related inquiry & 0.00 & 5.13 & 0.00 & 0.00 \\
Personal-related inquiry & 0.00 & 5.13 & 2.78 & 4.76 \\
None & 0.00 & 0.00 & 0.00 & 0.00 \\
\hline
\end{tabular}
\end{table*}
\begin{table*}[htbp]
\caption{Persuasion strategy prevalence for the movie theater dilemma by agent condition. $N$ denotes the total number of annotated agent messages per condition. Multiple strategies could be assigned to a single message; therefore, percentages may sum to more than 100\%.}
\Description{Percentage of agent messages containing each persuasion strategy. Multiple strategies may be assigned to a single message.}
\label{tab:strategies_movie}
\centering
\small
\renewcommand{\arraystretch}{1.15}
\begin{tabular}{lrrrr}
\hline
\textbf{Strategy} & \textbf{Base, $N=38$} & \textbf{Neutral, $N=35$} & \textbf{Aligned, $N=38$} & \textbf{Mismatched, $N=37$} \\
\hline
Logical appeal & 100.00 & 100.00 & 100.00 & 100.00 \\
Emotion appeal & 2.63 & 2.86 & 15.79 & 5.41 \\
Credibility appeal & 2.63 & 11.43 & 5.26 & 16.22 \\
Foot-in-the-door & 0.00 & 0.00 & 0.00 & 0.00 \\
Self-modeling & 0.00 & 8.57 & 0.00 & 2.70 \\
Personal story & 0.00 & 0.00 & 0.00 & 2.70 \\
Source-related inquiry & 0.00 & 2.86 & 0.00 & 0.00 \\
Task-related inquiry & 10.53 & 2.86 & 2.63 & 8.11 \\
Personal-related inquiry & 7.89 & 2.86 & 5.26 & 2.70 \\
None & 0.00 & 0.00 & 0.00 & 0.00 \\
\hline
\end{tabular}
\end{table*}
\newpage

\section{LLM Arguments per Dilemma}
To complement the exploratory conversation analysis in\Cref{sec:llm-assisted-labeling}, this section provides a more detailed overview of the arguments employed by LLM agents across the four moral dilemmas. Since agents were instructed to advocate against participants' initial choices, the arguments are reported separately for each initial position. The frequencies represent coded argument occurrences rather than numbers of conversations. Detailed distributions are provided in \cref{tab:coworker_arguments,tab:elder_arguments,tab:journal_arguments,tab:movie_arguments}.
The arguments differed by the participant’s initial position. In the co-worker dilemma, agents responding to participants who would accept the money argued against accepting it: they warned that paying the co-worker could reward laziness or normalize shortcuts (Base 2/6; Neutral 1/3; Mismatched 1/8) and that the co-worker would not learn the job (Base 2/6; Neutral 1/3). No Aligned counterarguments were recorded for this initial position. When participants refused the money, agents argued for accepting it, most often on the grounds of fairness and training. The claim that extra pay was fair appeared in 4/12 Base, 7/18 Neutral, 8/24 Aligned, and 6/22 Mismatched argument occurrences; the claim that the arrangement would improve the co-worker’s training appeared in 4/12, 7/18, 5/24, and 3/22, respectively.
The same split appeared in the other dilemmas. For participants who wanted to continue showing kindness to the rude patient, agents argued for setting firmer boundaries: they said tolerating rudeness reinforces abusive behavior (Base 3/8; Neutral 3/7; Aligned 6/10; Mismatched 2/6) and urged participants to protect their own well-being (3/8; 3/7; 4/10; 3/6). For participants who wanted to stop showing kindness, agents argued for continuing it, most often claiming that kindness could change the man’s behavior (4/9; 2/10; 1/7; 4/10).
In the journal dilemma, agents responding to participants who would read the journal argued against reading it, citing privacy, trust, and consent (Neutral 2/7; Aligned 1/2; Mismatched 1/3). In response to participants who would not read it, agents argued for reading it to check for danger or crisis (Base 7/20; Neutral 5/14; Aligned 6/17; Mismatched 5/20).
In the theater dilemma, agents responding to participants who would confront the teenagers argued against confrontation, citing possible violence (2/5; 3/7; 3/7; 5/15) and recommending that staff handle it (1/5; 3/7; 3/7; 5/15). For participants who would not confront them, agents argued for intervening, saying that stepping in would show the behavior was unacceptable (2/10 Neutral; 5/16 Aligned; 3/9 Mismatched) or that the situation could spiral out of control (5/16 Aligned; 3/9 Mismatched). These are coded argument occurrences, not counts of conversations.

\begin{table*}[htbp]
\caption{Most frequently occurring counterarguments in the co-worker dilemma, grouped by initial participant stance and agent condition. Frequencies represent the proportion of coded argument occurrences within each group.}
\Description{The three most frequent counterarguments
used by each LLM agent condition for both initial
stances in the co-worker dilemma.}
\label{tab:coworker_arguments}

\centering
\renewcommand{\arraystretch}{1.3}

\begin{tabular}{
    p{0.19\textwidth}
    p{0.13\textwidth}
    p{0.60\textwidth}
}
\hline
\textbf{Initial stance} &
\textbf{Agent} &
\textbf{Top three counterarguments
(occurrences/total arguments, \%)} \\
\hline

\textbf{Accept the money}
& Base
& 1. It rewards laziness and that shortcuts are acceptable in the company (2/6, 33.33\%) \newline
  2. The co-worker is not really learning his job (2/6, 33.33\%) \newline
  3. It creates a damaging culture in the company (1/6, 16.67\%) \\

& Neutral
& 1. It rewards laziness and that shortcuts are acceptable in the company (1/3, 33.33\%) \newline
  2. The co-worker is not really learning his job (1/3, 33.33\%) \newline
  3. You could be violating company policies (1/3, 33.33\%) \\

& Aligned
& Never occurred in our participant pool \\

& Mismatched
& 1. It rewards laziness and that shortcuts are acceptable in the company (1/8, 12.5\%) \newline
  2. You could be violating company policies (1/8, 12.5\%) \newline
  3. It can backfire if found out (1/8, 12.5\%) \\

\hline

\textbf{Refuse the money}
& Base
& 1. Being paid extra for more work is fair (4/12, 33.33\%) \newline
  2. It results in better training for the co-worker (4/12, 33.33\%) \newline
  3. It benefits the whole team (2/12, 16.67\%) \\

& Neutral
& 1. Being paid extra for more work is fair (7/18, 38.89\%) \newline
  2. It results in better training for the co-worker (7/18, 38.89\%) \newline
  3. It benefits the whole team (2/18, 11.11\%) \\

& Aligned
& 1. Being paid extra for more work is fair (8/24, 33.33\%) \newline
  2. It results in better training for the co-worker (5/24, 20.83\%) \newline
  3. It benefits the whole team (5/24, 20.83\%) \\

& Mismatched
& 1. Being paid extra for more work is fair (6/22, 27.27\%) \newline
  2. It benefits the whole team (5/22, 22.73\%) \newline
  3. It results in better training for the co-worker (3/22, 13.64\%) \\

\hline
\end{tabular}
\end{table*}

\begin{table*}[htbp]
\caption{Most frequently occurring counterarguments in the elder caregiver dilemma, grouped by initial participant stance and agent condition. Frequencies represent the proportion of coded argument occurrences within each group.}
\Description{The three most frequent counterarguments used by each LLM agent condition for both initial stances in the elder caregiver dilemma.}
\label{tab:elder_arguments}

\centering
\renewcommand{\arraystretch}{1.3}

\begin{tabular}{
    p{0.19\textwidth}
    p{0.13\textwidth}
    p{0.60\textwidth}
}
\hline
\textbf{Initial stance} &
\textbf{Agent} &
\textbf{Top three counterarguments
(occurrences/total arguments, \%)} \\
\hline

\textbf{Continue to show kindness}
& Base
& 1. Tolerating his rudeness only reinforces the abusive behavior  (3/8, 37.5\%) \newline
  2. You have to protect your own well-being (3/8, 37.5\%) \newline
  3. The behavior won't stop unless you are blunt (1/8, 12.5\%) \\

& Neutral
& 1. Tolerating his rudeness only reinforces the abusive behavior (3/7, 42.86\%) \newline
  2. You have to protect your own well-being (3/7, 42.86\%) \newline
  3. If you maintain this behavior, others in your close circle will disrespect you (1/7, 14.29\%) \\

& Aligned
& 1. Tolerating his rudeness only reinforces the abusive behavior (6/10, 60.0\%) \newline
  2. You have to protect your own well-being (4/10, 40.0\%) \\

& Mismatched
& 1. You have to protect your own well-being (3/6, 50.0\%) \newline
  2. Tolerating his rudeness only reinforces the abusive behavior (2/6, 33.33\%) \newline
  3. If the employer truly valued you, they'd let you set those boundaries (1/6, 16.67\%) \\

\hline

\textbf{Stop showing kindness}
& Base
& 1. By continuing to show kindness, the man will stop (4/9, 44.44\%) \newline
  2. By not showing kindness, you reinforce the idea that cruelty is acceptable (3/9, 33.33\%) \newline
  3. You can set boundaries while still being kind (1/9, 11.11\%) \\

& Neutral
& 1. By continuing to show kindness, the man will stop (2/10, 20.0\%) \newline
  2. By not showing kindness, you reinforce the idea that cruelty is acceptable (2/10, 20.0\%) \newline
  3. You have to treat every patient with dignity (1/10, 10.0\%) \\

& Aligned
& 1. The man might be struggling with fear of abandonment or loneliness (2/7, 28.57\%) \newline
  2. By continuing to show kindness, the man will stop (1/7, 14.29\%) \newline
  3. By not showing kindness, you reinforce the idea that cruelty is acceptable (1/7, 14.29\%) \\

& Mismatched
& 1. By continuing to show kindness, the man will stop (4/10, 40.0\%) \newline
  2. You have to stay patient to avoid fueling bitterness (3/10, 30.0\%) \newline
  3. The man might be struggling with fear of abandonment or loneliness (1/10, 10.0\%) \\

\hline
\end{tabular}
\end{table*}

\begin{table*}[htbp]
\caption{Most frequently occurring counterarguments in the journal reading dilemma, grouped by initial participant stance and agent condition. Frequencies represent the
proportion of coded argument occurrences within each group.}
\label{tab:journal_arguments}

\centering
\renewcommand{\arraystretch}{1.3}

\begin{tabular}{
    p{0.19\textwidth}
    p{0.13\textwidth}
    p{0.60\textwidth}
}
\hline
\textbf{Initial stance} &
\textbf{Agent} &
\textbf{Top three counterarguments
(occurrences/total arguments, \%)} \\
\hline

\textbf{Read the journal}
& Base
& Never occurred in our participant pool. \\

& Neutral
& 1. It crosses privacy, trust, and consent boundaries
 (2/7, 28.57\%) \newline
  2. You could misinterpret the wording and jump to false conclusions
 (1/7, 14.29\%) \newline
  3. You could still offer support without reading the journal
 (1/7, 14.29\%) \\

& Aligned
& 1. It crosses privacy, trust, and consent boundaries (1/2, 50.0\%) \newline
  2. You could still offer support without reading the journal (1/2, 50.0\%) \\

& Mismatched
& 1. It crosses privacy, trust, and consent boundaries (1/3, 33.33\%) \newline
  2. You could misinterpret the wording and jump to false conclusions (1/3, 33.33\%) \newline
  3. If the roles were reversed, would you feel safe sharing a flat with you (1/3, 33.33\%) \\

\hline

\textbf{Do not read the journal}
& Base
& 1. The odd behavior can be genuine danger, self-destructive, depression, crisis, etc.
 (7/20, 35.0\%) \newline
  2. The same story will not be told compared to the words in the journal
 (6/20, 30.0\%) \newline
  3. As the journal was open, it is a clear signal for help, and there is no intrusion into privacy
 (4/20, 20.0\%) \\

& Neutral
& 1. The odd behavior can be genuine danger, self-destructive, depression, crisis, etc. (5/14, 35.71\%) \newline
  2. The same story will not be told compared to the words in the journal (5/14, 35.71\%) \newline
  3. As the journal was open, it is a clear signal for help, and there is no intrusion into privacy (4/14, 28.57\%) \\

& Aligned
& 1. The odd behavior can be genuine danger, self-destructive, depression, crisis, etc. (6/17, 35.29\%) \newline
  2. Prevents harm to the person and ourselves.
 (4/17, 23.53\%) \newline
  3. The same story will not be told compared to the words in the journal (4/17, 23.53\%) \\

& Mismatched
& 1. Prevents harm to the person and ourselves. (6/20, 30.0\%) \newline
  2. The odd behavior can be genuine danger, self-destructive, depression, crisis, etc. (5/20, 25.0\%) \newline
  3. The same story will not be told compared to the words in the journal (4/20, 20.0\%) \\

\hline
\end{tabular}

\Description{The three most frequent counterarguments
used by each LLM agent condition for both initial
stances in the journal reading dilemma.}

\end{table*}

\begin{table*}[htbp]
\caption{Most frequently occurring counterarguments in the movie theater dilemma, grouped by initial participant stance and agent condition. Frequencies represent the
proportion of coded argument occurrences within each group.}
\label{tab:movie_arguments}

\centering
\renewcommand{\arraystretch}{1.3}

\begin{tabular}{
    p{0.19\textwidth}
    p{0.13\textwidth}
    p{0.60\textwidth}
}
\hline
\textbf{Initial stance} &
\textbf{Agent} &
\textbf{Top three counterarguments
(occurrences/total arguments, \%)} \\
\hline

\textbf{Confront teenagers}
& Base
& 1. It can become violent, even physical (2/5, 40.0\%) \newline
  2. We should trust the staff to handle the situation (1/5, 20.0\%) \newline
  3. By confronting the teenagers, you will disrupt the others in the theatre (1/5, 20.0\%) \\

& Neutral
& 1. It can become violent, even physical (3/7, 42.86\%) \newline
  2. We should trust the staff to handle the situation (3/7, 42.86\%) \newline
  3. It is not snitching to alarm the staff (1/7, 14.29\%) \\

& Aligned
& 1. It can become violent, even physical (3/7, 42.86\%) \newline
  2. We should trust the staff to handle the situation (3/7, 42.86\%) \newline
  3. Respect the teenagers by dealing with them without publicly shaming them (1/7, 14.29\%) \\

& Mismatched
& 1. It can become violent, even physical (5/15, 33.33\%) \newline
  2. We should trust the staff to handle the situation (5/15, 33.33\%) \newline
  3. By confronting the teenagers, you will disrupt the others in the theatre (3/15, 20.0\%) \\

\hline

\textbf{Do not confront}
& Base
& 1. Mentioning it calmly but confidently will help (4/14, 28.57\%) \newline
  2. Once you confront, the rest of the theatre will step up (3/14, 21.43\%) \newline
  3. Waiting for staff makes the ruined experience longer (2/14, 14.29\%) \\

& Neutral
& 1. Waiting for staff makes the ruined experience longer (5/10, 50.0\%) \newline
  2. By stepping up, you show that the behavior is not acceptable (2/10, 20.0\%) \newline
  3. By not stepping up, you make the experience worse for everyone (2/10, 20.0\%) \\

& Aligned
& 1. By stepping up, you show that the behavior is not acceptable (5/16, 31.25\%) \newline
  2. It is better to step up now before the situation spirals out of control (5/16, 31.25\%) \newline
  3. Waiting for staff makes the ruined experience longer (4/16, 25.0\%) \\

& Mismatched
& 1. By stepping up, you show that the behavior is not acceptable (3/9, 33.33\%) \newline
  2. It is better to step up now before the situation spirals out of control (3/9, 33.33\%) \newline
  3. Waiting for staff makes the ruined experience longer (1/9, 11.11\%) \\

\hline
\end{tabular}

\Description{The three most frequent counterarguments
used by each LLM agent condition for both initial
stances in the movie theatre dilemma.}

\end{table*}

\newpage
\newpage

\section{LLM Argument Repetition}
This section provides the distribution of argument repetition across the four agent conditions (\Cref{tab:repetition}). Conversations were manually categorized as showing no repetition, some repetition (revisiting arguments while introducing new information), or substantial repetition (repeating arguments with little or no new information). The table reports the frequency of each category by agent condition, complementing the exploratory conversation analysis in~\Cref{sec:llm-assisted-labeling}.

\begin{table*}[htbp]
\caption{Distribution of manually annotated repetition levels
across agent conditions and dilemmas. Values represent the
percentage of conversations assigned each repetition level.
Each dilemma--agent combination contains eight conversations
($N=8$). Categories are mutually exclusive.}
\label{tab:repetition}
\centering
\renewcommand{\arraystretch}{1.2}
\setlength{\tabcolsep}{8pt}
\begin{tabular}{llrrrr}
\hline
\textbf{Dilemma} &
\textbf{Repetition} &
\textbf{Base} &
\textbf{Neutral} &
\textbf{Aligned} &
\textbf{Mismatched} \\
\hline

\textbf{Elderly man}
& None        & 12.5 &  0.0 &  0.0 &  0.0 \\
& Some        & 37.5 & 37.5 & 25.0 & 50.0 \\
& Substantial & 50.0 & 62.5 & 75.0 & 50.0 \\
\hline

\textbf{Movie theater}
& None        &  0.0 &  0.0 &  0.0 &  0.0 \\
& Some        & 50.0 & 75.0 & 62.5 & 37.5 \\
& Substantial & 50.0 & 25.0 & 37.5 & 62.5 \\
\hline

\textbf{Journal}
& None        &  0.0 & 12.5 &  0.0 &  0.0 \\
& Some        & 62.5 & 50.0 & 12.5 & 25.0 \\
& Substantial & 37.5 & 37.5 & 87.5 & 75.0 \\
\hline

\textbf{Co-worker training}
& None        & 12.5 &  0.0 &  0.0 & 25.0 \\
& Some        & 25.0 & 50.0 & 25.0 & 50.0 \\
& Substantial & 62.5 & 50.0 & 75.0 & 25.0 \\
\hline
\end{tabular}
\Description{Percentage of conversations manually classified
as exhibiting no repetition, some repetition, or substantial
repetition, grouped by dilemma and agent personality condition.}
\end{table*}

\newpage
\section{User Frustration}
To complement the exploratory conversation analysis in~\Cref{sec:llm-assisted-labeling}, this section presents the distribution of observed participant frustration across the four agent conditions. Table~\ref{tab:explicit_frustration_conditions} reports the number of conversations in which participants expressed frustration during their interactions with the LLM agents.
\begin{table*}[htbp]
\caption{Explicit User Frustration by Dilemma and Agent Condition}
\Description{Table counting conversations in which participants expressed explicit frustration, by moral dilemma (rows) and agent condition (columns: Base, Neutral, Aligned, Mismatched), with a total column. Each dilemma and condition combination contains eight conversations. Frustration was most frequent in the Journal dilemma (8 conversations), followed by Co-worker training (7), Movie theater (6) and Elderly man (1). Across dilemmas, the personality-conditioned agents account for most cases: Mismatched 9, Aligned 8, Neutral 3 and Base 2, 22 in total.}
\label{tab:explicit_frustration_conditions}
\begin{tabular}{lrrrrr}
\toprule
 \textbf{Dilemma}& \textbf{Base} & \textbf{Neutral} & \textbf{Aligned} & \textbf{Mismatched} & \textbf{Total} \\

\midrule
\textbf{Journal} & 1 & 0 & 3 & 4 & 8 \\
\textbf{Co-worker training} & 1 & 2 & 2 & 2 & 7 \\
\textbf{Movie theater} & 0 & 1 & 3 & 2 & 6 \\
\textbf{Elderly man} & 0 & 0 & 0 & 1 & 1 \\
\bottomrule
\end{tabular}
\end{table*}

\end{document}